\documentclass[conference]{IEEEtran}
\IEEEoverridecommandlockouts

\usepackage{cite}
\usepackage{amsmath,amssymb,amsfonts}
\usepackage{algorithmic}
\usepackage{graphicx}
\usepackage{textcomp}
\usepackage{xcolor}
\usepackage{subcaption}
\usepackage{multirow}
\usepackage{hyperref}
\usepackage{ulem}
\usepackage{booktabs}
\def\BibTeX{{\rm B\kern-.05em{\sc i\kern-.025em b}\kern-.08em
    T\kern-.1667em\lower.7ex\hbox{E}\kern-.125emX}}
\begin{document}

\title{Performance Analysis of HPC applications on the Aurora Supercomputer: Exploring the Impact of HBM-Enabled Intel Xeon Max CPUs
}



\author{
    \IEEEauthorblockN{Huda Ibeid\IEEEauthorrefmark{1}, Vikram Narayana\IEEEauthorrefmark{1}, Jeongnim Kim\IEEEauthorrefmark{1}, Anthony Nguyen\IEEEauthorrefmark{1}, Vitali Morozov\IEEEauthorrefmark{4}, Ye Luo\IEEEauthorrefmark{4}}
    \IEEEauthorblockA{\IEEEauthorrefmark{1}Intel Corporation \{huda.ibeid, vikram.narayana, jeongnim.kim, anthony.d.nguyen\}@intel.com}
    \IEEEauthorblockA{\IEEEauthorrefmark{4}Argonne National Laboratory, Lemont, IL, USA \{morozov,yeluo\}@anl.gov}
}

\maketitle

\begin{abstract}
The Aurora supercomputer is an exascale-class system designed to tackle some of the most demanding computational workloads. Equipped with both High Bandwidth Memory (HBM) and DDR memory, it provides unique trade-offs in performance, latency, and capacity. This paper presents a comprehensive analysis of the memory systems on the Aurora supercomputer, with a focus on evaluating the trade-offs between HBM and DDR memory systems. We explore how different memory configurations, including memory modes (Flat and Cache) and clustering modes (Quad and SNC4), influence key system performance metrics such as memory bandwidth, latency, CPU-GPU PCIe bandwidth, and MPI communication bandwidth. Additionally, we examine the performance of three representative HPC applications — HACC, QMCPACK, and BFS — each illustrating the impact of memory configurations on performance. By using microbenchmarks and application-level analysis, we provide insights into how to select the optimal memory system and configuration to maximize performance based on the application characteristics. The findings presented in this paper offer guidance for users of the Aurora system and similar exascale systems.
\end{abstract}


\section{Introduction}
The Aurora supercomputer at the Argonne Leadership Computing Facility is an exascale-class system capable of performing over a quintillion calculations per second (10\(^{18}\))~\cite{aurora}. To support these massive computational demands, Aurora is equipped with 12.2 PB of CPU memory and 8.16 PB of GPU HBM memory. The CPU memory consists of 10.9 PB of DDR5 and 1.36 PB of HBM memory, providing a unique blend of performance characteristics that are suited to a wide range of applications.

Memory systems such as DDR and HBM differ significantly in bandwidth, latency, and capacity, each influencing system performance in different ways. HBM offers much higher bandwidth, making it ideal for memory-intensive applications, while DDR, with its lower latency and larger capacity, is more advantageous for workloads with lower memory access demands. Understanding the trade-offs between these memory systems is critical for optimizing performance on systems like Aurora.

This paper presents a comprehensive analysis of the memory systems on the Aurora supercomputer, focusing on the trade-offs between HBM and DDR memory systems. We explore how different memory configurations impact both system performance metrics and application performance. Specifically, we evaluate memory modes (Flat and Cache) and clustering modes (Quad and SNC4), emphasizing their effects on key metrics such as memory bandwidth, latency, CPU-GPU PCIe bandwidth, and MPI communication bandwidth. Additionally, we examine the performance of three representative HPC applications — HACC, QMCPACK, and BFS — each demonstrating distinct trade-offs between HBM and DDR memory systems and configurations. While this paper focuses on memory and system performance, GPU optimizations are outside its scope; for a detailed discussion of GPU performance on the Aurora supercomputer, we refer readers to~\cite{applencourt2024ponte}.

While several studies have benchmarked HBM memory performance and compared it to DDR~\cite{mccalpin2023bandwidth,fukazawa2024,siegmann2024}, to the best of our knowledge, this is the first paper to provide a comprehensive analysis of the memory systems and memory and clustering modes of HBM-enabled CPUs on the Aurora supercomputer. We focus not only on the performance of the memory systems but also on their implications for system metrics and HPC applications.

We make the following contributions:
\begin{itemize}
\item We use microbenchmarks to evaluate the impact of memory systems and their configurations on system performance. Specifically, we employ the STREAM benchmark and Intel Memory Latency Checker to measure memory bandwidth and latency, a custom benchmark for CPU-GPU PCIe bandwidth evaluation, and the \texttt{osu\_mbw\_mr} test from the OSU benchmark suite to analyze MPI bandwidth.
\item We evaluate the performance of HPC applications, including HACC, QMCPACK, and BFS, under different memory systems and configurations, highlighting the trade-offs between memory systems for various workloads.
\item We provide guidance on how to select memory systems and configurations to optimize performance based on application characteristics.
\end{itemize}

\begin{figure*}[t]
\centerline{\includegraphics[width=0.75\textwidth]{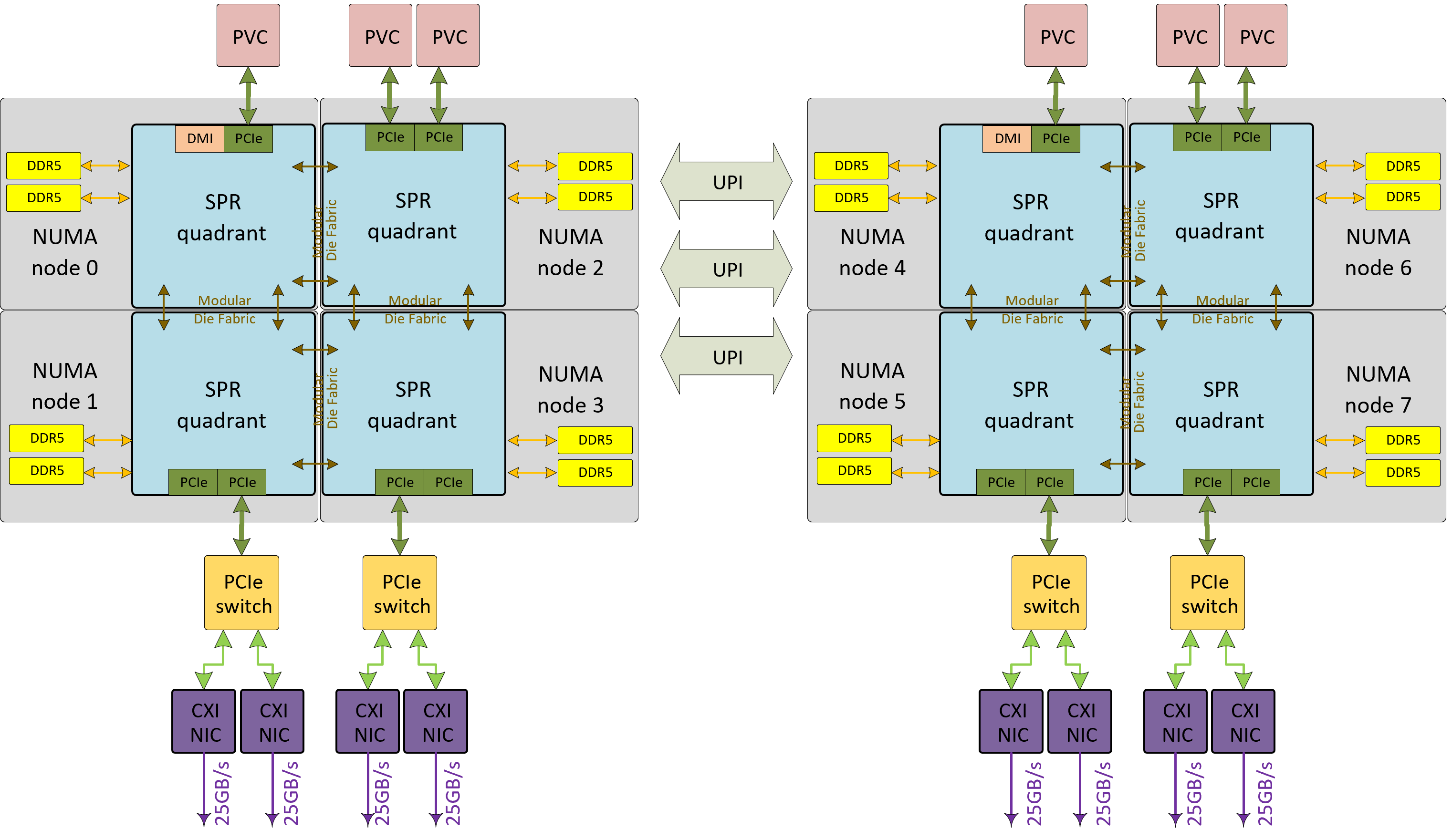}}
\caption{Block diagram of the Aurora Exascale Compute Blade in SNC4 mode, where each CPU socket is divided into four quadrants. The blade features six Ponte Vecchio (PVC) GPUs and a dual-socket Sapphire Rapids (SPR) CPU interconnected via three UPI links. Each CPU socket connects to three GPUs and four NICs through x16 PCIe Gen5 links.}
\label{fig:blade_snc4}
\end{figure*}

The paper is organized as follows: Section~\ref{sec:background} provides background on the Aurora compute blade architecture, including an overview of its memory systems and the various memory and clustering modes. Section~\ref{sec:setup} presents the experimental setup and configuration. Section~\ref{sec:microbenchmarks} discusses the impact of memory systems and configurations on system performance metrics, such as memory bandwidth, latency, CPU-GPU PCIe bandwidth, and MPI bandwidth. Section~\ref{sec:hpc_apps} presents the HPC applications used in our study, along with their performance results, highlighting the impact of memory systems on application performance. Section~\ref{sec:relatedwork} presents related work. Finally, Section~\ref{sec:discussion} provides the discussion and conclusion, summarizing our findings.

\section{Background} \label{sec:background}

\subsection{Aurora Supercomputer: Compute Blade Architecture}
\label{subsec:aurora_blade}

The Aurora supercomputer features a hierarchical architecture consisting of 166 compute racks. Each rack contains eight chassis, with each chassis housing eight Exascale Compute Blades (ECBs) and four network switches that interconnect the compute blades.

Figure~\ref{fig:blade_snc4} presents the block diagram of the Aurora Exascale Compute Blade. Each compute blade is equipped with six Intel Data Center GPU Max Series units, referred to by their architectural codename, Ponte Vecchio (PVC), in the figure. Each GPU comprises two tiles and eight high-bandwidth memory (HBM) stacks. The blade also includes a dual-socket Intel Xeon Max 9470C CPU, formerly codenamed Sapphire Rapids (SPR), with each socket featuring four HBM stacks, access to eight DDR5 memory channels, and a last-level cache (LLC) of 105 MiB.

Each compute blade is equipped with 768 GB of HBM for the GPUs, along with 128 GB of HBM and 1,024 GB of DDR5 memory for the CPUs. The GPU memory achieves a peak bandwidth of 3.27 TB/s per GPU, while each CPU socket delivers a peak HBM bandwidth of 1.43 TB/s and a peak DDR5 bandwidth of 0.28 TB/s.

The two CPU sockets are interconnected via three Intel Ultra Path Interconnect (UPI) links. Each socket connects to three GPUs and four network interface cards (NICs) through x16 PCIe Gen5 links. Each NIC provides approximately 25 GB/s of bandwidth, yielding a total peak aggregate bandwidth of 200 GB/s per blade.

The compute blade delivers a peak performance of 186.2 TFLOPS in double-precision floating-point operations, with peak GPU performance of 29.7 TFLOPS and peak CPU performance of 8 TFLOPS.

Across all 166 compute racks, the Aurora system provides a total memory capacity of 8.16 PB of GPU HBM, 1.36 PB of CPU HBM, and 10.9 PB of DDR5 memory. Each rack comprises 64 compute blades and 32 switch blades, yielding 49.1 TB of GPU HBM, 8.2 TB of CPU HBM, and 64 TB of DDR5 memory per rack.

\subsection{Overview of HBM Modes in Intel Xeon Max CPUs}

\begin{figure}[htbp]
    \centering
    \begin{subfigure}[b]{0.45\linewidth}
        \centering
        \includegraphics[width=\linewidth]{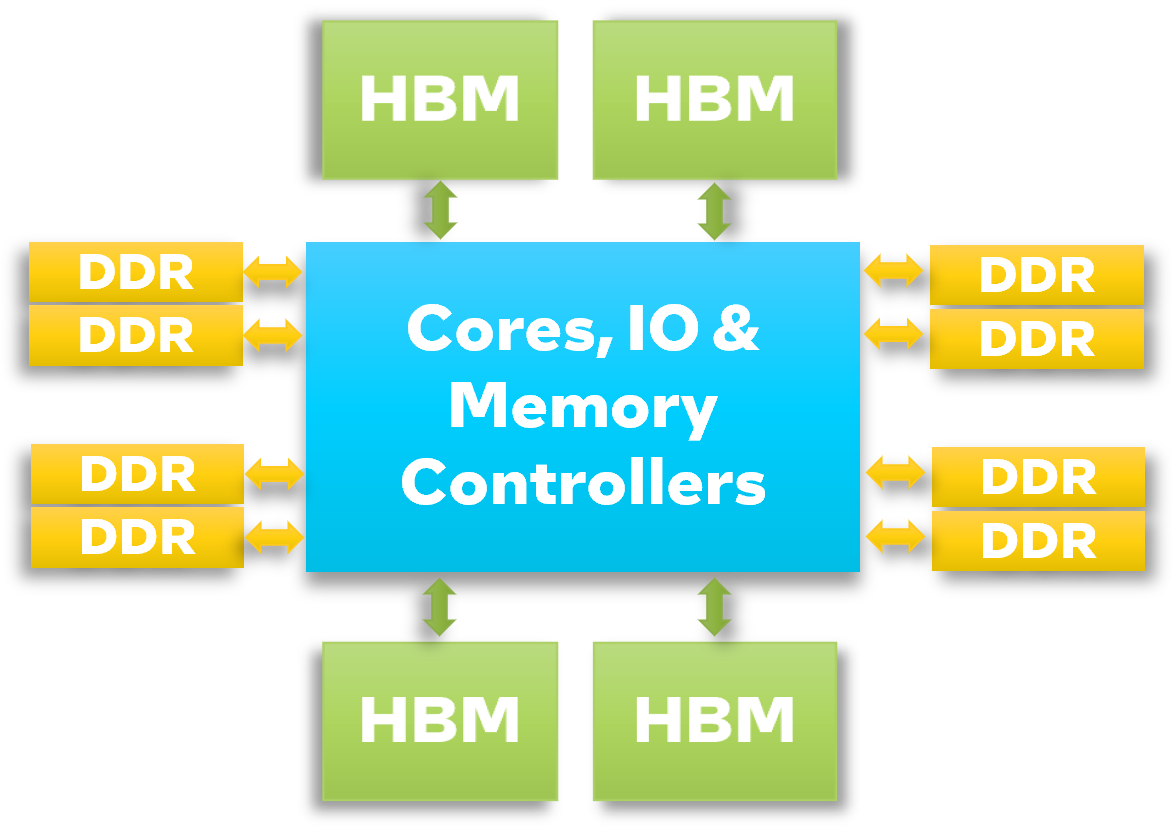}
        \caption{Flat, Quad}
        \label{fig:f_quad}
    \end{subfigure}
    \hfill
    \begin{subfigure}[b]{0.45\linewidth}
        \centering
        \includegraphics[width=\linewidth]{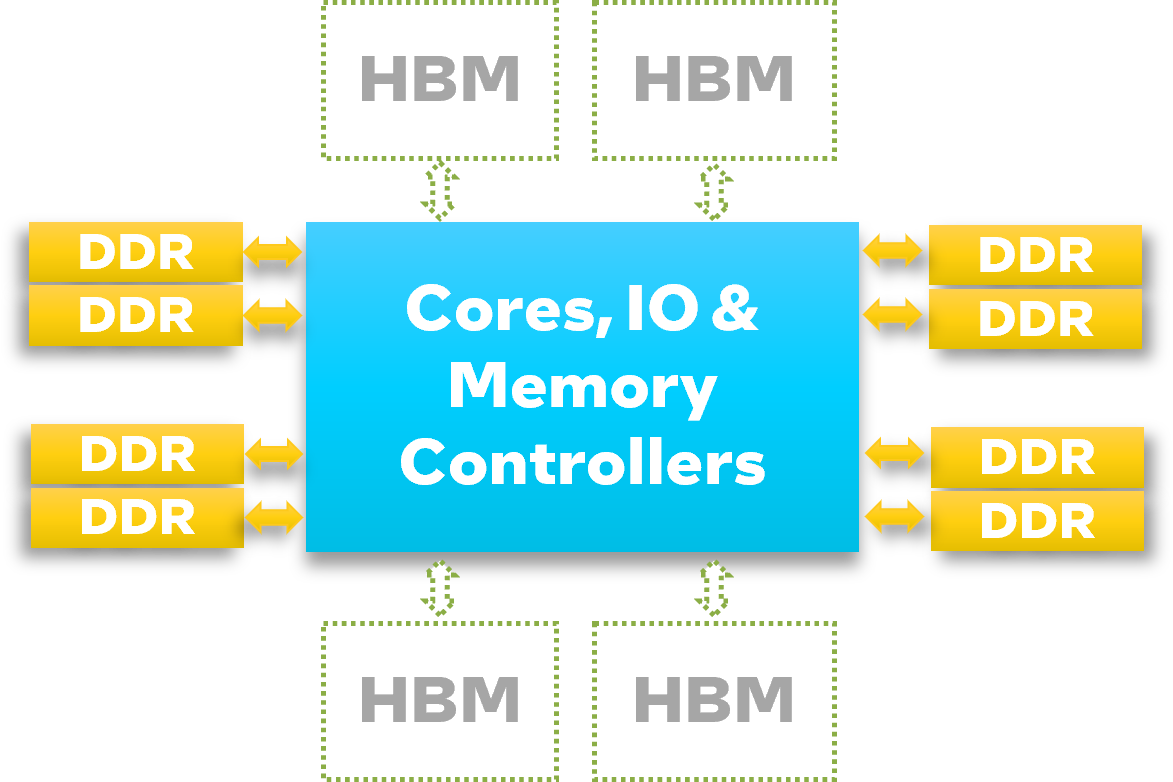}
        \caption{Cache, Quad}
        \label{fig:c_quad}
    \end{subfigure}
    \vskip\baselineskip
    \begin{subfigure}[b]{0.45\linewidth}
        \centering
        \includegraphics[width=\linewidth]{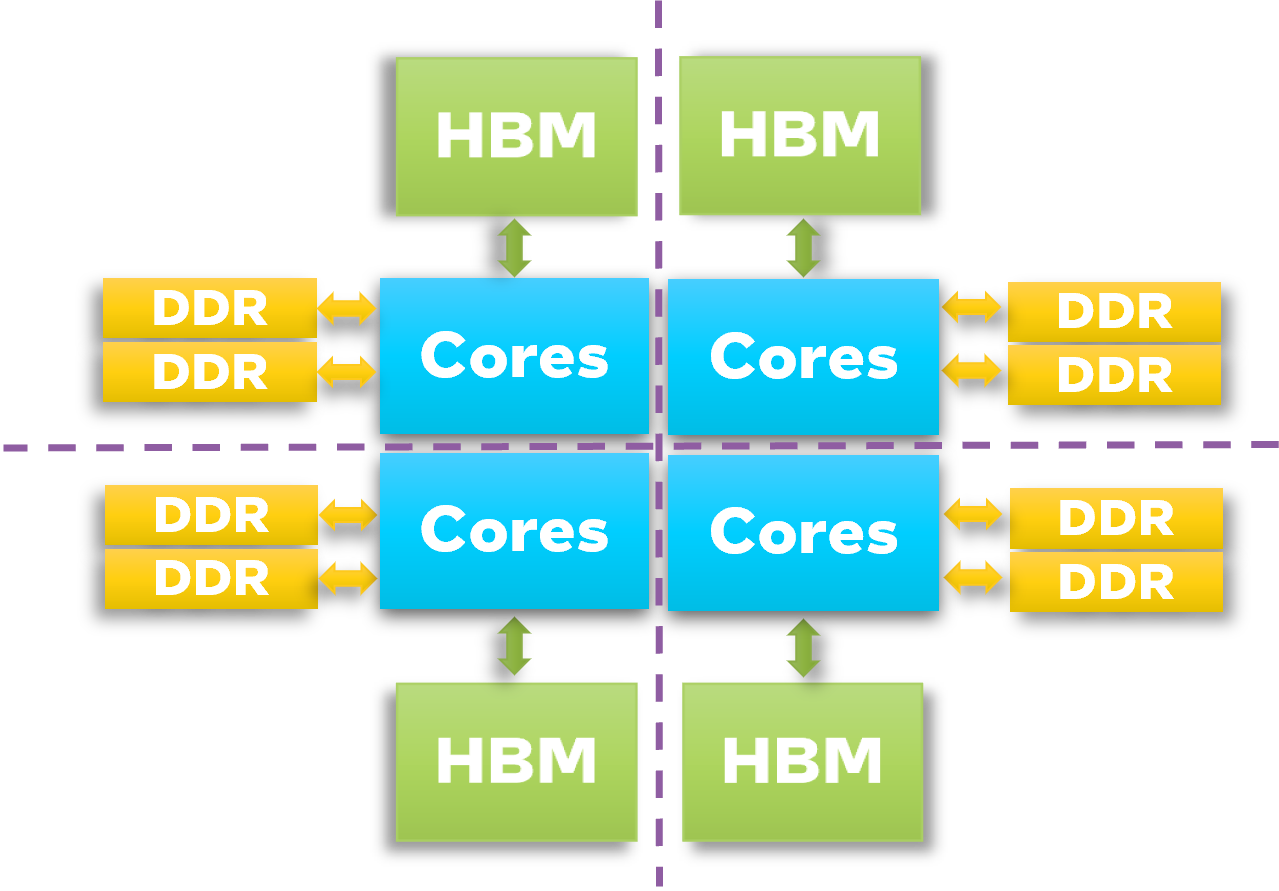}
        \caption{Flat, SNC4}
        \label{fig:f_snc4}
    \end{subfigure}
    \hfill
    \begin{subfigure}[b]{0.45\linewidth}
        \centering
        \includegraphics[width=\linewidth]{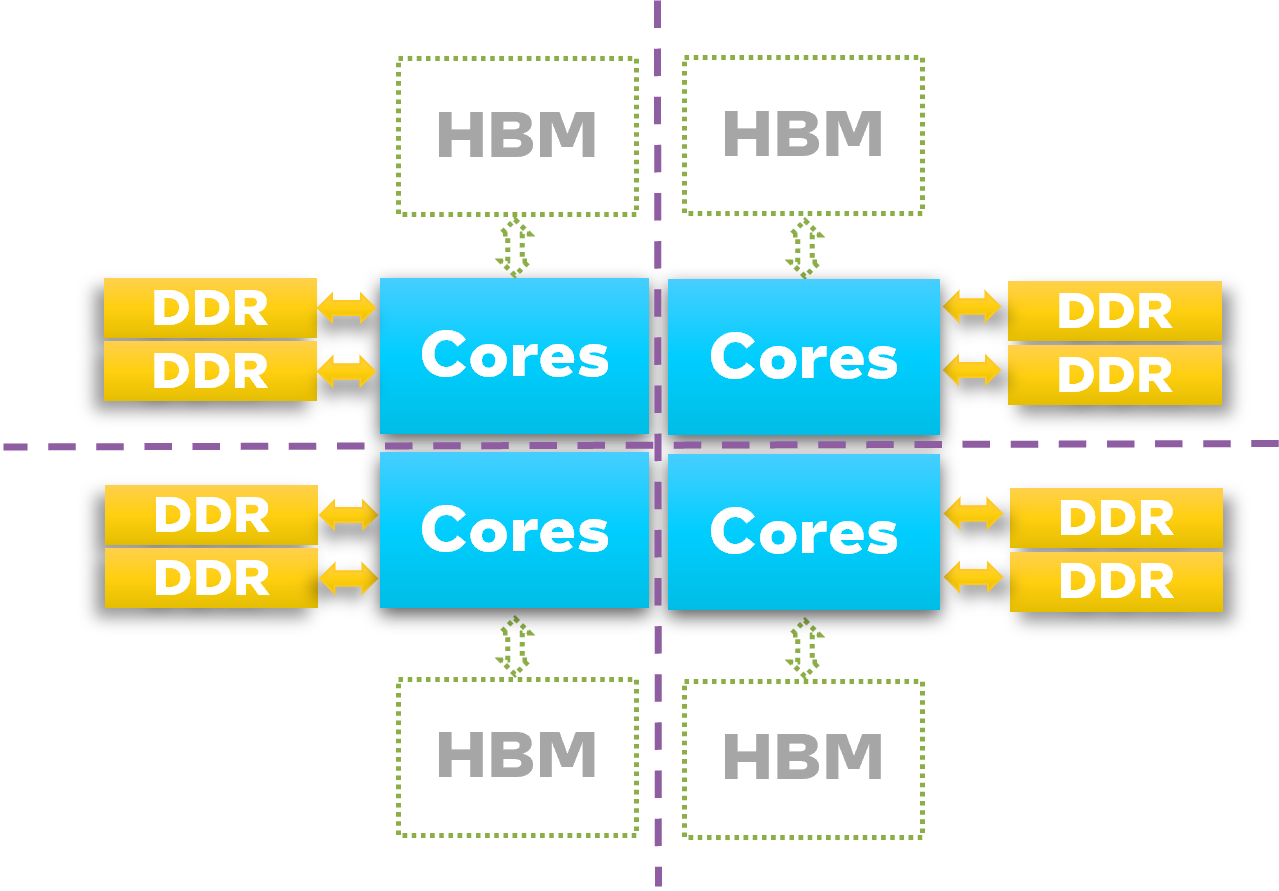}
        \caption{Cache, SNC4}
        \label{fig:c_snc4}
    \end{subfigure}
    \caption{Memory modes x Clustering modes~\cite{intel_xeon_max_guide}.}
    \label{fig:hbm_modes}
\end{figure}

The Intel Xeon Max CPU with HBM offers two memory modes and two clustering modes. These modes are orthogonal, resulting in four possible configuration combinations, as summarized in Figure~\ref{fig:hbm_modes}. This section provides a hardware-level overview of each mode and discusses strategies for achieving optimal performance with HBM; additional details are available in~\cite{intel_xeon_max_guide}.

\subsubsection{Memory Modes}

The CPU exposes HBM to the operating system and applications through two distinct memory modes: Flat and Cache, described below.

\paragraph{Flat Mode} In Flat mode, HBM and DDR are presented as separate, addressable memory spaces (NUMA nodes), with HBM and DDR each mapped to different NUMA nodes. To effectively utilize HBM in this mode, users can employ NUMA-aware tools (e.g., \texttt{numactl}) or libraries, such as \texttt{libnuma}, for memory management and optimization. The best performance is typically achieved when the application’s memory footprint fits within the HBM capacity. Therefore, optimizing memory usage to remain within this limit is crucial. Techniques such as distributing workloads across multiple sockets using MPI or employing shared-memory approaches like OpenMP within a socket can help minimize memory consumption.

\paragraph{Cache Mode} In Cache mode, only the DDR address space is visible to software, with HBM functioning as a transparent cache for DDR. Applications can leverage this mode without requiring modifications.

The HBM cache operates as a direct-mapped cache, which may require specific configurations to mitigate conflict misses. Cache mode is particularly effective for applications with memory footprints exceeding the HBM capacity, especially when page shuffling is enabled to reduce performance variability. Additionally, enabling Fake NUMA in Cache mode is recommended for applications with memory requirements that are within the HBM capacity. Fake NUMA~\cite{fake_numa} partitions memory into multiple NUMA nodes, each representing a contiguous block of memory. By partitioning DDR memory into Fake NUMA nodes with sizes equal to the HBM capacity and directing memory access to a single Fake NUMA node, memory access is restricted to the HBM size, which helps mitigate conflict misses in the direct-mapped cache.

\subsubsection{Clustering Modes}

These modes define how the CPU is partitioned into multiple addressable memory spaces (NUMA nodes), with each partition grouping specific cores and their associated memory. This enables cores to access both HBM and DDR memory within their partition, improving memory locality, which reduces latency and increases bandwidth. Intel Xeon Max CPUs support two clustering modes: Quad and SNC4 (Sub-NUMA Clustering-4), described below.

\paragraph{Quad Mode} In Quad mode, the CPU presents a single address space (NUMA node) to the operating system and applications. This mode is ideal for applications that are not NUMA-aware and benefit from shared access to large data structures across all cores, such as OpenMP workloads that utilize all cores of a CPU.

\paragraph{SNC4 Mode} In SNC4 mode, the CPU is partitioned into four quadrants, each containing one or more NUMA nodes, depending on the memory configuration. Compared to Quad mode, SNC4 offers higher bandwidth and lower latency, making it well-suited for NUMA-aware applications, such as those using MPI or hybrid MPI+OpenMP models. This mode is recommended for applications that can be efficiently decomposed across quadrants, thereby optimizing memory locality and maximizing performance.

\section{Experimental Setup and Configuration} \label{sec:setup}

For single-node performance evaluation, we assess all combinations of memory and clustering modes on Borealis, a testbed for the Aurora system. The evaluated modes include Quad-Flat/HBM, SNC4-Flat/HBM, Quad-Flat/DDR, SNC4-Flat/DDR, Quad-Cache, and SNC4-Cache.

For multinode scaling performance evaluation, experiments are conducted directly on the Aurora system. Since Aurora is currently configured to operate in Quad-Flat mode, and this configuration cannot be modified at present, we compare the performance of Quad-Flat/HBM with Quad-Flat/DDR.

The HPC applications used in this study are configured with problem sizes and setups consistent with those employed in the acceptance testing of the Aurora system. Unless otherwise specified, all experiments are repeated three times, and the mean values are reported.

\section{System Performance Evaluation Using Micro-Benchmarks} \label{sec:microbenchmarks}

In this section, we evaluate system performance by examining key metrics, including memory bandwidth and latency, CPU-GPU PCIe bandwidth, and MPI bandwidth. We compare the performance of HBM and DDR memory systems, and also analyze the effects of different memory modes (Flat and Cache) and clustering configurations (Quad and SNC4) on these metrics. Through targeted micro-benchmarks, we assess how these memory systems and configurations influence the overall performance of the Aurora system.

\subsection{Memory Bandwidth and Latency}

We use the STREAM Triad benchmark~\cite{mccalpin1995memory} to evaluate memory bandwidth across different memory and clustering modes for both HBM and DDR. As shown in Table~\ref{tab:stream_bw}, HBM in SNC4-Flat mode delivers the highest STREAM bandwidth, with SNC4 consistently outperforming Quad mode across all configurations. The next highest bandwidth is observed in HBM in Quad-Flat, followed by SNC4-Cache, Quad-Cache, DDR in SNC4-Flat, and finally DDR in Quad-Flat. Moreover, the sustained memory bandwidth of HBM is significantly higher than that of DDR. However, the observed increase in sustained bandwidth is smaller than the theoretical peak bandwidth difference, primarily due to insufficient memory concurrency, as discussed in~\cite{mccalpin2023bandwidth}.
\begin{table}[htbp]
    \centering
    \caption{Single-socket STREAM Triad Bandwidth (GB/s).}
    \renewcommand{\arraystretch}{1.3}
    \begin{tabular}{lccc}
        \toprule
        \multirow{2}{*}{\textbf{Clustering Mode}} & \multicolumn{2}{c}{\textbf{Flat Mode}} & \multirow{2}{*}{\textbf{Cache Mode}} \\ \cline{2-3}
         & \textbf{DDR} & \textbf{HBM} & \\ 
        \midrule
        \textbf{Quad} (Full socket) & 235 & 680 & 560 \\ 
        \textbf{SNC4} (Full socket) & 245 & 840 & 575 \\ 
        \textbf{SNC4} (CPU quadrant) & 60 & 210 & 150 \\ 
        \bottomrule
    \end{tabular}
    \label{tab:stream_bw}
\end{table}

Additionally, we measure the memory bandwidth and access latency for both HBM and DDR across different memory and clustering modes using the Intel Memory Latency Checker (v3.11b). Each test is run 10 times to account for variability, and we report the mean and range of observed latencies. Both \textit{idle} and \textit{loaded} latencies are captured using the \texttt{--idle\_latency} and \texttt{--loaded\_latency} options, respectively.

As shown in Table~\ref{tab:idle_latency}, DDR exhibits lower idle latency across all memory and clustering modes compared to HBM. In addition, SNC4 clustering mode consistently achieves more than 10\% lower idle latency than Quad mode across all memory configurations.
\begin{table}[htbp]
    \centering
    \caption{Single-socket idle memory access latency (ns), presented as mean (range).}
    \renewcommand{\arraystretch}{1.3}
    \begin{tabular}{lccc}
        \toprule
        \multirow{2}{*}{\textbf{Clustering Mode}} & \multicolumn{2}{c}{\textbf{Flat Mode}} & \multirow{2}{*}{\textbf{Cache Mode}} \\ \cline{2-3}
         & \textbf{DDR} & \textbf{HBM} & \\ 
        \midrule
        \multirow{2}{*}{\textbf{Quad}} & 112.9 & 134.9 & 136.3 \\ 
             & (112.7--113.1) & (134.7--135.3) & (136.0--136.9) \\ 
        \multirow{2}{*}{\textbf{SNC4}} & 96.3 & 121.1 & 121.6 \\ 
             & (96.1--97.5) & (120.9--121.2) & (121.5--121.7) \\ 
        \bottomrule
    \end{tabular}
    \label{tab:idle_latency}
\end{table}

\begin{figure}[htbp]
\centerline{\includegraphics[width=0.75\linewidth]{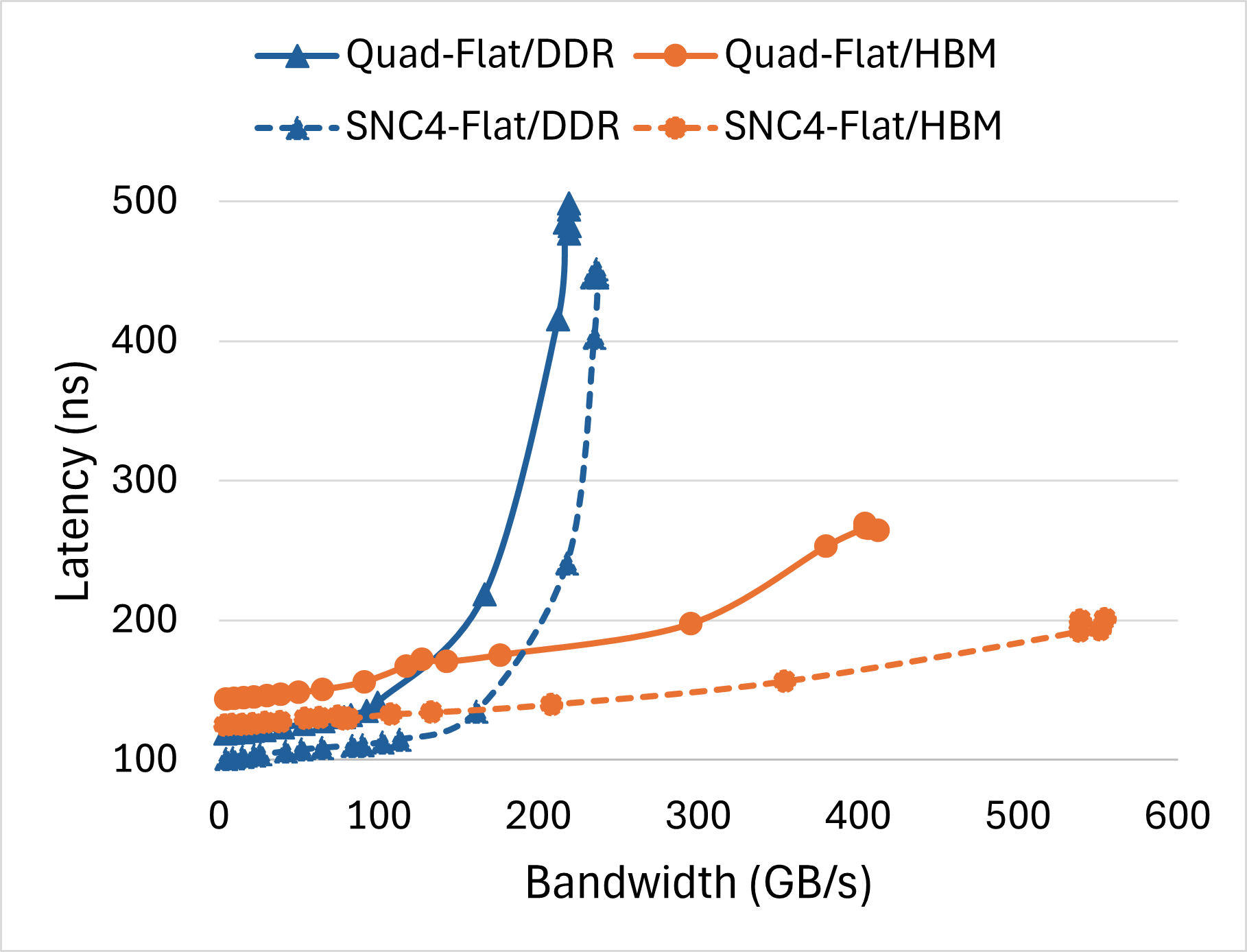}}
\caption{Single-socket memory bandwidth and loaded latency.}
\label{fig:loaded_latency}
\end{figure}

Figure~\ref{fig:loaded_latency} illustrates the relationship between memory bandwidth and loaded latency for both HBM and DDR in Flat memory mode, based on experiments conducted using the \texttt{--loaded\_latency -W10} configuration. The \texttt{-W10} flag specifies a 2:1 read to non-temporal write ratio.

On the left side of Figure~\ref{fig:loaded_latency}, where memory access loads are lower, DDR demonstrates lower latency than HBM across all clustering modes, making it well-suited for scenarios involving small, frequent memory accesses. However, as the access load increases, DDR memory latency exceeds that of HBM, indicating reduced efficiency under heavier memory demands. This trend highlights that while DDR is more effective for handling quick, small requests, HBM is better suited for applications requiring high bandwidth, where large data transfers and sustained throughput are essential.

Figure~\ref{fig:loaded_latency} also compares the performance of the Quad and SNC4 clustering modes. As shown, SNC4 reduces latency and achieves higher memory bandwidth for both HBM and DDR, due to its improved locality.

\subsection{CPU-GPU PCIe Bandwidth}

Each data transfer between the CPU and a PCIe device requires a corresponding memory read or write operation. In systems where a CPU socket supports multiple x16 PCIe Gen5 links, such as the Aurora blade described in Section~\ref{subsec:aurora_blade}, memory bandwidth can become a bottleneck, limiting data transfer rates on the PCIe links. HBM memory helps mitigate this bottleneck when multiple PCIe links are active simultaneously. We demonstrate this through experiments involving data transfers between a CPU socket and the GPUs connected to it on the Aurora blade.

We use a microbenchmark in which each process allocates a 256 MiB buffer in both CPU memory and GPU memory, and performs unidirectional or bidirectional data transfers between the two buffers. Timing these transfers measures the bandwidth achieved on the respective x16 PCIe Gen5 link. On the Aurora blade, maximum PCIe bandwidth is achieved when both compute tiles of each GPU are utilized, due to the presence of copy engines on each tile. Therefore, the microbenchmark uses two processes per GPU, one per tile. The aggregate bandwidth achieved by the two processes is reported as the PCIe bandwidth per GPU.

\begin{table}[htbp]
    \centering
    \caption{Theoretical maximum CPU memory bandwidth demand (GB/s) for PCIe transfers.}
    \renewcommand{\arraystretch}{1.3}
    \begin{tabular}{lcc}
        \toprule
        & \textbf{1 GPU} & \textbf{3 GPUs} \\ 
        \midrule
        \textbf{Unidirectional} & 55 & 165 \\ 
        \textbf{Bidirectional}  & 76 & 228 \\ 
        \bottomrule
    \end{tabular}
    \label{tab:pcie_mem_bw}
\end{table}
For the Aurora blade, Table~\ref{tab:pcie_mem_bw} shows the expected demand on CPU memory bandwidth for PCIe transfers. The one-GPU memory bandwidth demands represent the PCIe bandwidths measured under best-case scenarios, while the three-GPUs theoretical memory bandwidth demands are simply three times the values from the first column. Comparing Table~\ref{tab:pcie_mem_bw} with the STREAM bandwidths summarized in Table~\ref{tab:stream_bw}, it can be seen that when all three GPUs are active, the PCIe bandwidth demand approaches the achievable DDR bandwidth. Empirically, we observe that PCIe transfers are unable to fully utilize the memory bandwidths reported by the STREAM benchmark. Therefore, when PCIe demand exceeds approximately 80\% of the memory bandwidth benchmarks, a degradation in PCIe bandwidth can be expected, making DDR memory bandwidth a limiting factor for PCIe performance.

\begin{figure}[htbp]
    \centering
    \begin{subfigure}[b]{0.75\linewidth}
        \centering
        \includegraphics[width=\linewidth]{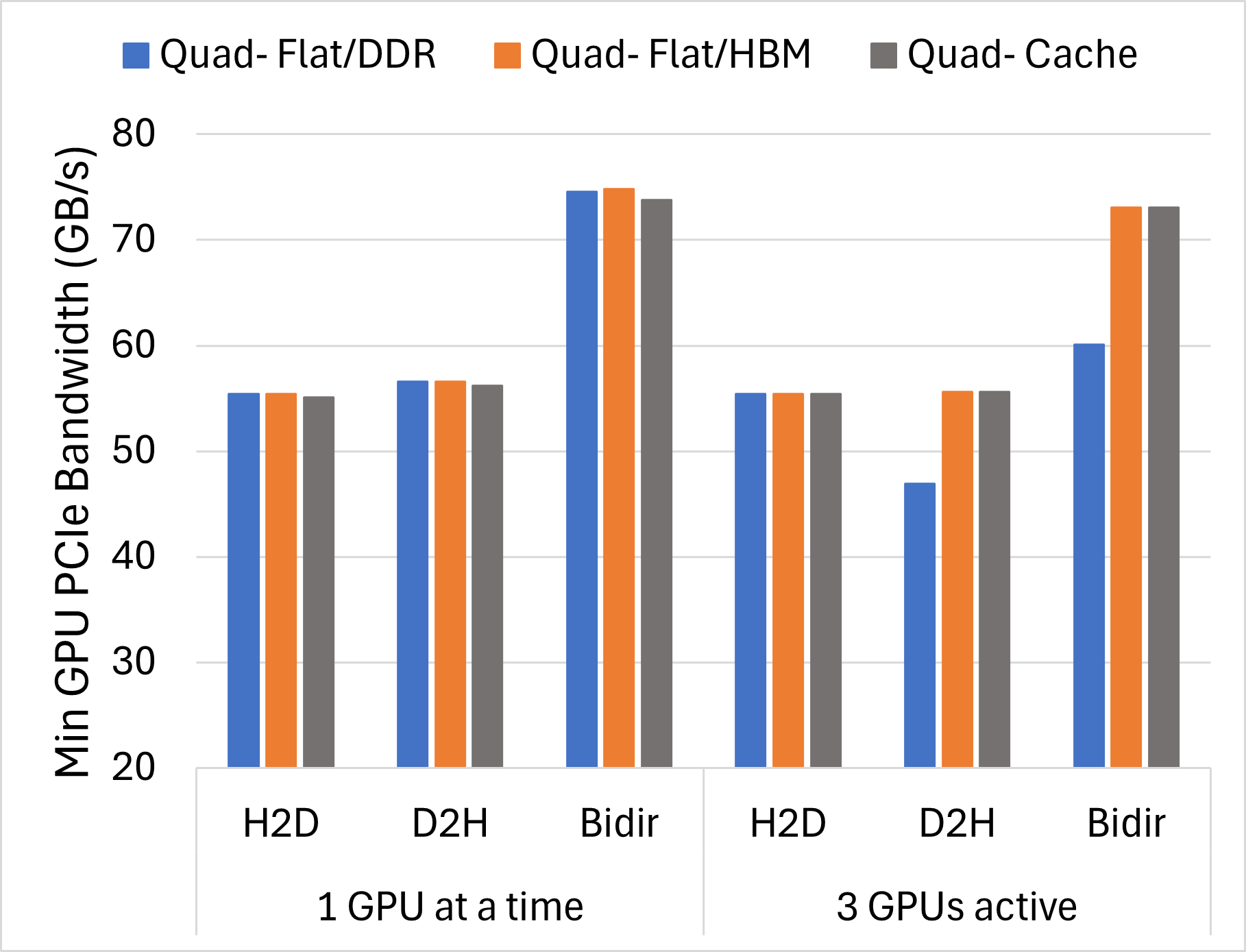}
        \caption{Quad mode}
        \label{fig:pcie_quad}
    \end{subfigure}
    \begin{subfigure}[b]{0.75\linewidth}
        \centering
        \includegraphics[width=\linewidth]{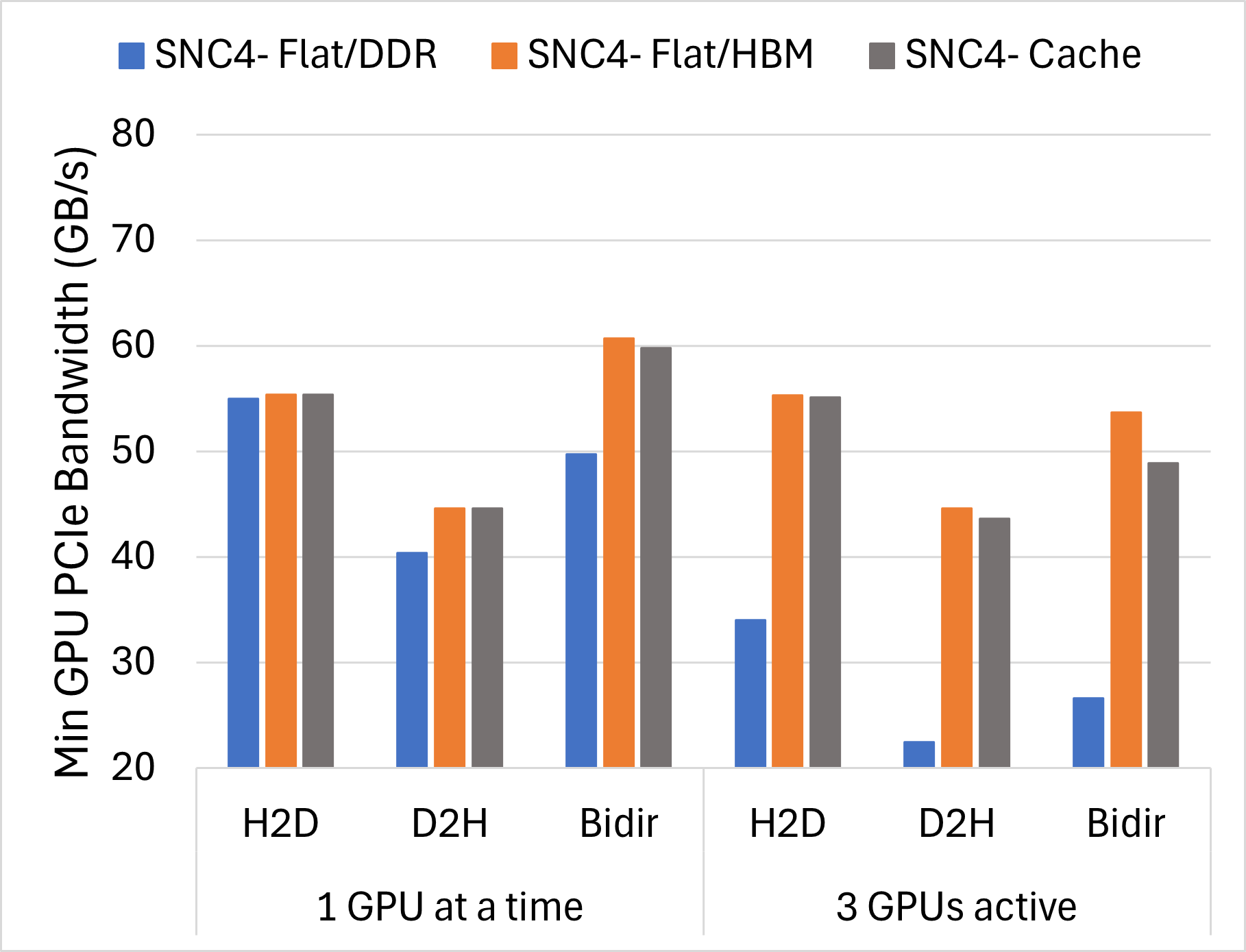}
        \caption{SNC4 mode}
        \label{fig:pcie_snc4}
    \end{subfigure}
    \caption{CPU-GPU PCIe bandwidth across different memory and clustering modes.}
    \label{fig:pcie_bw}
\end{figure}

Figure~\ref{fig:pcie_quad} summarizes the per-GPU PCIe bandwidth for unidirectional host-to-device (H2D), device-to-host (D2H), and bidirectional (Bidir) data transfers in Quad clustering mode. As expected, no differences are observed between memory modes when only one GPU is active. However, simultaneous data transfers across all three GPUs reveal the limitations of DDR memory, particularly for D2H and Bidir transfers, while H2D bandwidth remains unaffected. D2H transfers, which require memory writes, are constrained by the lower write bandwidth of DDR. A custom write-only STREAM benchmark achieves 197 GB/s in Quad-Flat/DDR mode, approximately 17\% lower than the STREAM Triad 235 GB/s (Table~\ref{tab:stream_bw}). This may explain why D2H transfers achieve only 47 GB/s per GPU, compared to 55 GB/s for H2D transfers. Similarly, bidirectional bandwidth in Flat/DDR mode is reduced to 60 GB/s per GPU when all three GPUs are active, compared to 75 GB/s in single-GPU tests. These results highlight that in Quad clustering mode, using HBM, whether in Cache or Flat mode, is preferable for achieving higher PCIe bandwidth.

Figure~\ref{fig:pcie_snc4} presents the PCIe bandwidth results for the SNC4 clustering mode. In these tests, the two processes associated with each GPU are pinned to CPU cores within the nearest NUMA quadrant, according to Figure~\ref{fig:blade_snc4}. In SNC4 mode, each quadrant is allocated only one-fourth of the total memory bandwidth, as shown in the ``CPU quadrant'' row of Table~\ref{tab:stream_bw}. Consequently, even the single-GPU D2H bandwidth is affected when using DDR. While Flat/HBM and Cache modes improve the D2H bandwidth, the performance remains below the best single-GPU D2H results observed in Quad clustering mode. This suggests a potential limitation for D2H and bidirectional transfers in SNC4 mode.

\begin{figure}[htbp]
    \centering
    \includegraphics[width=0.75\linewidth]{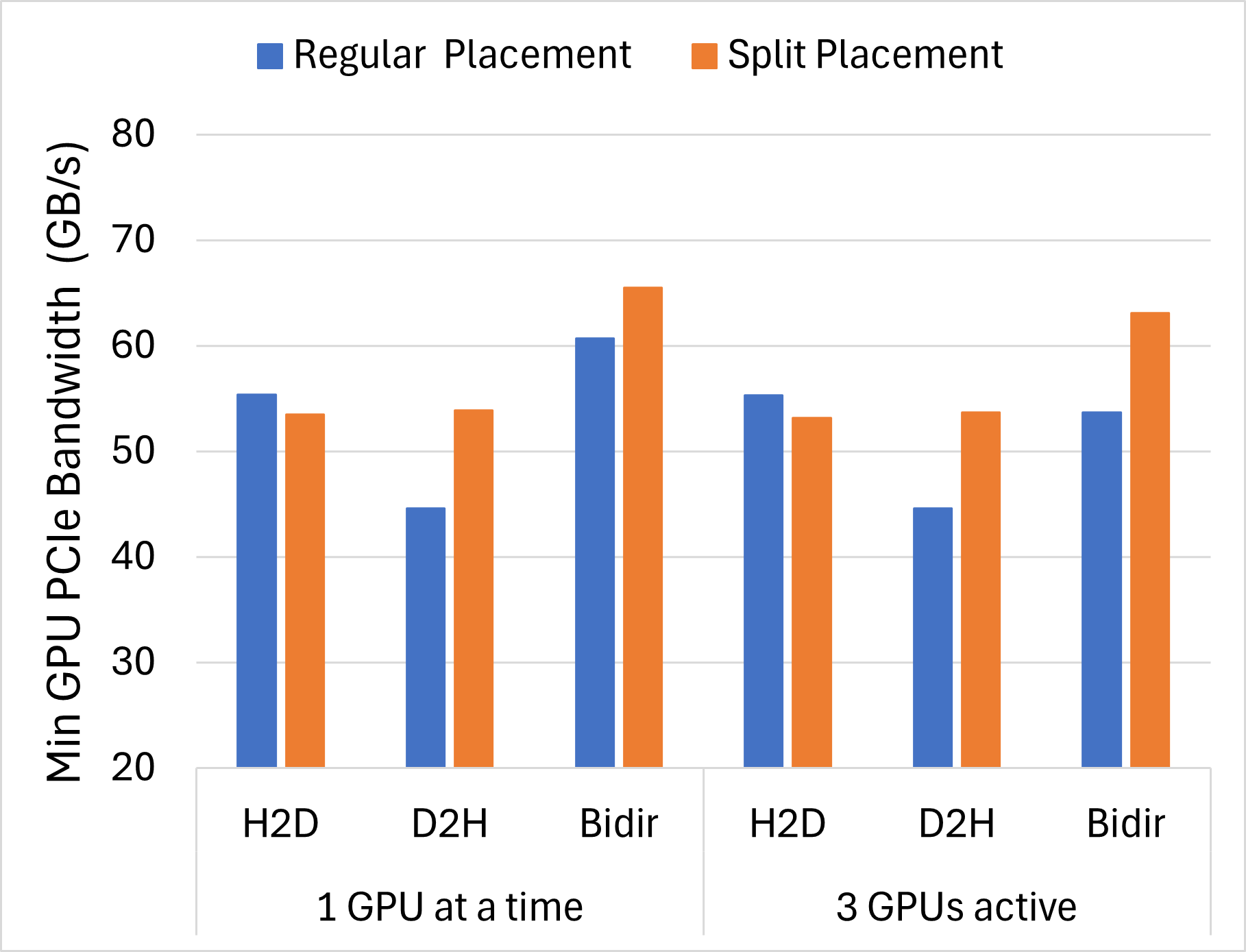}
    \caption{GPU PCIe bandwidth in SNC4-Flat/HBM mode with two different process placement strategies.}
    \label{fig:snc4_pinning_pcie_bw}
\end{figure}

Figure~\ref{fig:pcie_snc4} further shows that in SNC4 mode, the best H2D performance for three-GPU tests is achieved in Flat/HBM and Cache modes, as expected. However, D2H and bidirectional bandwidth remain lower than the corresponding performance in Quad mode, likely due to inherent limitations. Additionally, the degraded bidirectional bandwidth is partly attributed to two GPUs being attached to the same CPU quadrant (see Figure~\ref{fig:blade_snc4}). Consequently, the processes associated with these GPUs share the memory bandwidth within the quadrant, creating a bottleneck for PCIe bidirectional bandwidth. Both of the aforementioned issues can be partially addressed by pinning processes across different CPU quadrants, if the workload allows. Table~\ref{tab:snc4_pinning_options} illustrates an example. In our tests, two processes are associated with each GPU. In regular placement, both processes are mapped to the CPU quadrant nearest to the GPU, whereas in split placement, the two processes are mapped to separate quadrants. The table lists the CPU core IDs that the processes are pinned to, along with the quadrant IDs, which follow the same numbering scheme as the NUMA nodes in Figure~\ref{fig:blade_snc4}. Split placement leverages memory associated with two quadrants while ensuring that one of the processes remains in a quadrant close to the GPU. The results obtained in SNC4-Flat/HBM mode with these two process placement options are shown in Figure~\ref{fig:snc4_pinning_pcie_bw}.
\begin{table}[htbp]
    \centering
    \caption{Process pinning options for the two processes associated with each GPU in SNC4 mode.}
    \renewcommand{\arraystretch}{1.3}
    \begin{tabular}{llccc}
        \cline{3-5}
        & & \textbf{GPU 0} & \textbf{GPU 1} & \textbf{GPU 2}\\ 
        \midrule
        \multirow{2}{*}{\textbf{Regular Placement}} & \textbf{CPU core ID}                  & 1, 2 & 27, 28 & 29, 30 \\
        \cline{2-5}
        & \textbf{CPU quadrant} & 0, 0 & 2, 2  & 2, 2 \\
        \midrule
        \multirow{2}{*}{\textbf{Split Placement}} & \textbf{CPU core ID} & 1, 14 & 27, 40 & 28, 41 \\
        \cline{2-5}
        & \textbf{CPU quadrant} & 0, 1 & 2, 3 & 2, 3 \\
        \bottomrule
    \end{tabular}
    \label{tab:snc4_pinning_options}
\end{table}

Firstly, with split process placement, single-GPU D2H and bidirectional (Bidir) bandwidth improve by 21\% and 8\%, respectively, indicating that this strategy mitigates some limitations affecting regular placement, while H2D bandwidth experiences only a marginal degradation. When three GPUs are active simultaneously, D2H and Bidir bandwidth improve by 20\% and 18\%, respectively, over regular placement. The improvement in D2H bandwidth primarily stems from the single-GPU performance gains. In contrast, the larger improvement in Bidir bandwidth suggests additional benefits from leveraging memory bandwidth across two CPU quadrants. This was further validated  through experiments with only two active GPUs. When the selected GPUs are connected to different CPU quadrants, the improvement from split process pinning aligns with the single-GPU results. However, when both GPUs are connected to the same CPU quadrant, Bidir bandwidth benefits significantly more from split process pinning, matching the three-GPU scenario. Despite these gains, split process placement experiences a performance penalty when accessing a GPU that is not local to the CPU quadrant, and the resulting bandwidth still falls short of the performance achieved in Quad-Flat/HBM mode (Figure~\ref{fig:pcie_quad}). In summary, workloads heavily reliant on PCIe bandwidth should prioritize using Quad clustering mode and ensure memory allocations target HBM for optimal performance.

\subsection{MPI Bandwidth}
\label{sec:mpi_communication}

To evaluate the impact of different memory modes on MPI bandwidth, we use the \texttt{osu\_mbw\_mr} test from the OSU benchmark suite~\cite{osubenchmark}. The benchmark is configured with the parameters \texttt{-i 25 -m 268435456}, where \texttt{-i 25} specifies 25 iterations per run and \texttt{-m 268435456} sets the maximum message size to 256 MiB. The test is conducted across two nodes within the same chassis.

\begin{figure}[htbp]
\centerline{\includegraphics[width=0.75\linewidth]{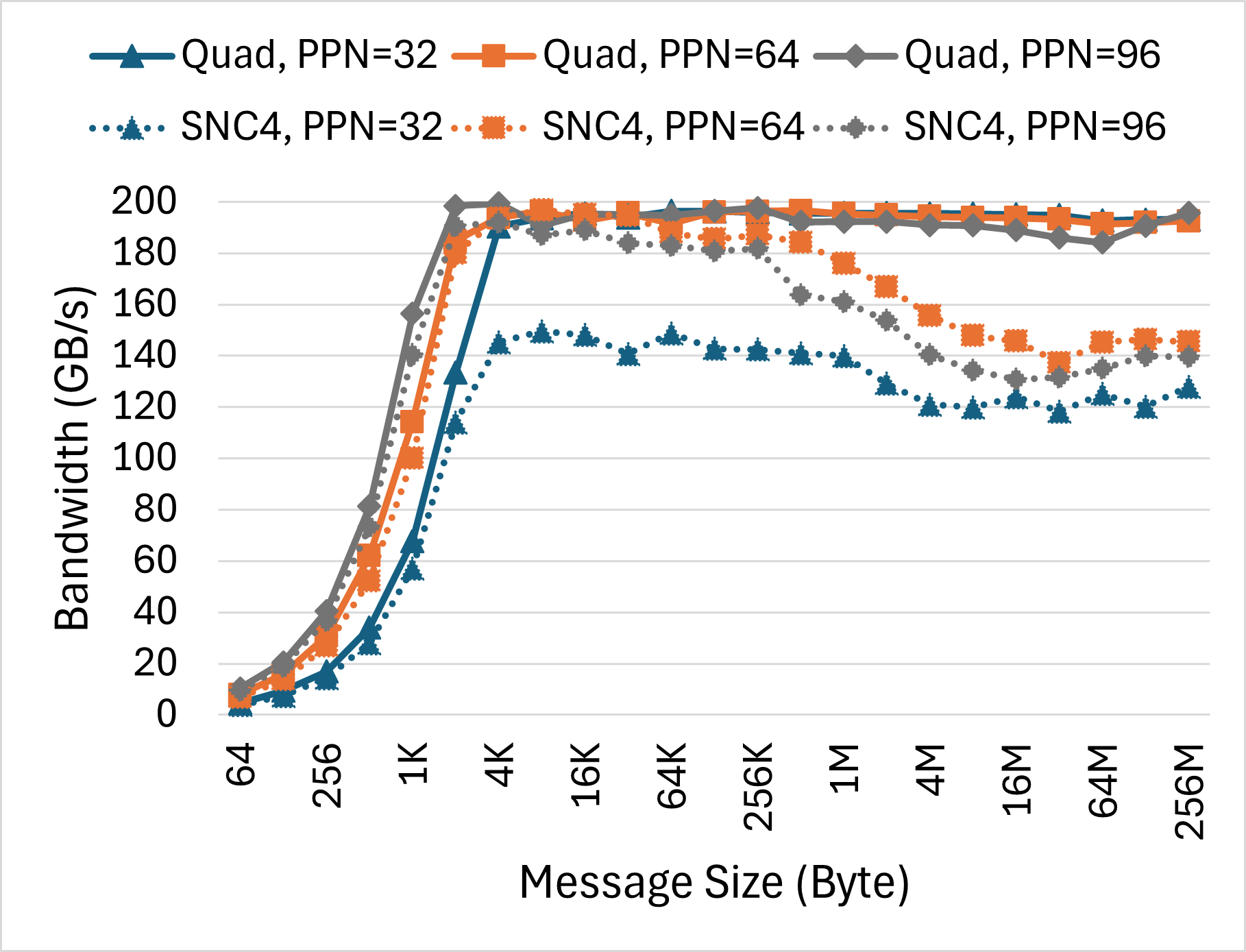}}
\caption{Comparison of MPI bandwidth between Quad-Flat/DDR and SNC4-Flat/DDR memory modes.}
\label{fig:osu_bw}
\end{figure}

Figure~\ref{fig:osu_bw} compares MPI bandwidth between the Quad-Flat/DDR and SNC4-Flat/DDR memory modes. The tests are performed with varying numbers of processes per node (\texttt{PPN}), evenly distributed across NUMA nodes.

The results in Figure~\ref{fig:osu_bw} show that Quad clustering mode sustains consistent peak bandwidth across various \texttt{PPN} configurations (32, 64, 96). In contrast, SNC4 mode experiences significant performance degradation, particularly at lower \texttt{PPN} values, with reductions ranging from 28\% to 38\% for the largest message size. This decline in SNC4 performance is attributed to asymmetric NIC connectivity. As shown in Figure~\ref{fig:blade_snc4}, in SNC4 mode, Quadrants 1 and 3 in each socket are directly connected to two NICs each, while Quadrants 0 and 2 rely on inter-quadrant communication. Here, the quadrant IDs follow the same numbering scheme as the NUMA nodes shown in Figure~\ref{fig:blade_snc4}.

In Quad clustering mode, MPI processes are assigned to NICs on a socket in a round-robin manner. In SNC4 mode, processes in Quadrants 1 and 3, which are directly connected to two NICs each, are assigned to the locally connected NICs in a round-robin fashion. For Quadrants 0 and 2, which lack direct NIC connectivity, processes are distributed across all NICs on the socket in a round-robin manner. This leads to inter-quadrant communication in SNC4 mode, causing congestion on the inter-quadrant links, which increases network latency and reduces available bandwidth.

\begin{figure}[htbp]
\centerline{\includegraphics[width=0.75\linewidth]{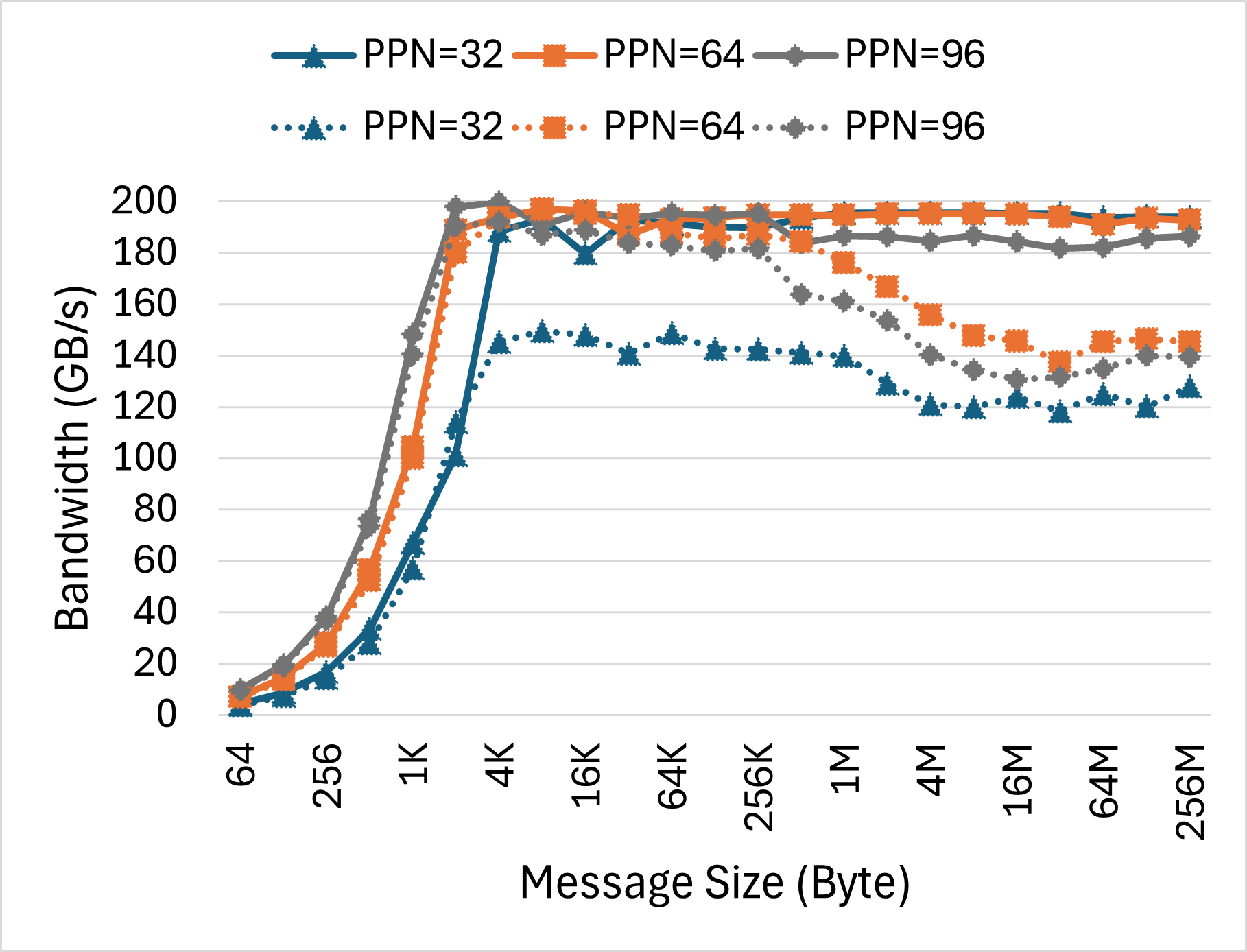}}
\caption{MPI bandwidth comparison in SNC4-Flat/DDR mode, highlighting the impact of Quadrant-to-NIC 1-to-1 mapping. Solid lines represent 1-to-1 mapping, while dashed lines represent the scenario without it.}
\label{fig:osu_bw_mapped}
\end{figure}

This issue is mitigated in the latest Aurora MPICH implementation, which, in SNC4 clustering mode, adopts a 1-to-1 mapping between Quadrants and NICs. In this approach, all MPI processes within a given Quadrant communicate exclusively through the NIC assigned to that Quadrant. This 1-to-1 mapping significantly improves bandwidth in SNC4 clustering mode, bringing performance in line with that of Quad mode, as shown in Figure~\ref{fig:osu_bw_mapped}.

A limitation of this method is that if a Quadrant has no MPI processes assigned, the corresponding NIC remains unused, potentially leading to underutilization of network resources. To address this, the Aurora MPICH introduces the MPIR\_CVAR\_CH4\_OFI\_PREF\_NIC environment variable. This variable enables explicit NIC selection for each MPI process and can be configured on a per-process basis in multi-NIC systems, allowing users to specify a preferred NIC for each process to optimize communication.

\section{Performance Evaluation of HPC Applications} \label{sec:hpc_apps}

To evaluate the impact of memory systems and their configurations on HPC applications, we analyze three representative workloads: HACC, QMCPACK, and BFS. Each application highlights different trade-offs between HBM and DDR memory systems.
\begin{itemize}
\item \textbf{HACC}, a cosmology simulation application, demonstrates the importance of memory bandwidth. It leverages the higher data transfer rates of HBM to maximize computational throughput.
\item \textbf{QMCPACK}, a quantum Monte Carlo simulation application, is sensitive to memory latency, showcasing the benefits of the lower latency of DDR in accelerating computations.
\item \textbf{BFS}, a graph traversal algorithm, highlights the significance of memory capacity, where the larger memory footprint of DDR offers performance advantages in certain scenarios.
\end{itemize}

The experiments are conducted using the \texttt{Intel oneAPI DPC++/C++ Compiler 2025} and \texttt{MKL 2025}. All applications are compiled with the \texttt{-O3 -ffast-math} optimization flags to maximize performance.

In the following subsections, we present the performance results for each application, analyzing how the memory systems and their configurations affect application performance and efficiency.

\subsection{HACC}

Hardware-Accelerated Cosmology Code (HACC) is a high-performance simulation framework designed to efficiently solve large-scale cosmological problems~\cite{habib2016hacc}. It employs advanced computational techniques, including the Recursive Coordinate Bisection (RCB) tree for particle interactions and domain decomposition. HACC is widely used for simulating dark matter, galaxy formation, and large-scale structure evolution in cosmology.

HACC computation consists of three distinct phases. The short-force evaluation kernel is compute-intensive with regular stride-one memory accesses, making it well-suited for vectorization and threading. The tree-walk phase involves highly irregular indirect memory accesses, frequent branching, and a high number of integer operations. Finally, the 3D FFT phase relies on point-to-point communication. 

We run HACC with the parameters \texttt{-w -R -N 512}, where \texttt{-w} applies the white noise initializer, \texttt{-R} enables the RCB monopole tree, and \texttt{-N 512} sets the maximum number of particles per RCB tree leaf. The grid size (\(n_g\)) and particle count (\(n_p\)) are set equal (\(n_p = n_g\)), ensuring a consistent particle density across the simulation domain, which is defined as a cube. Additionally, we configure the system to use \texttt{PPN=96}, specifying 96 processes per node.

\subsubsection{Single-Node Performance}

\begin{figure}[htbp]
\centerline{\includegraphics[width=0.75\linewidth]{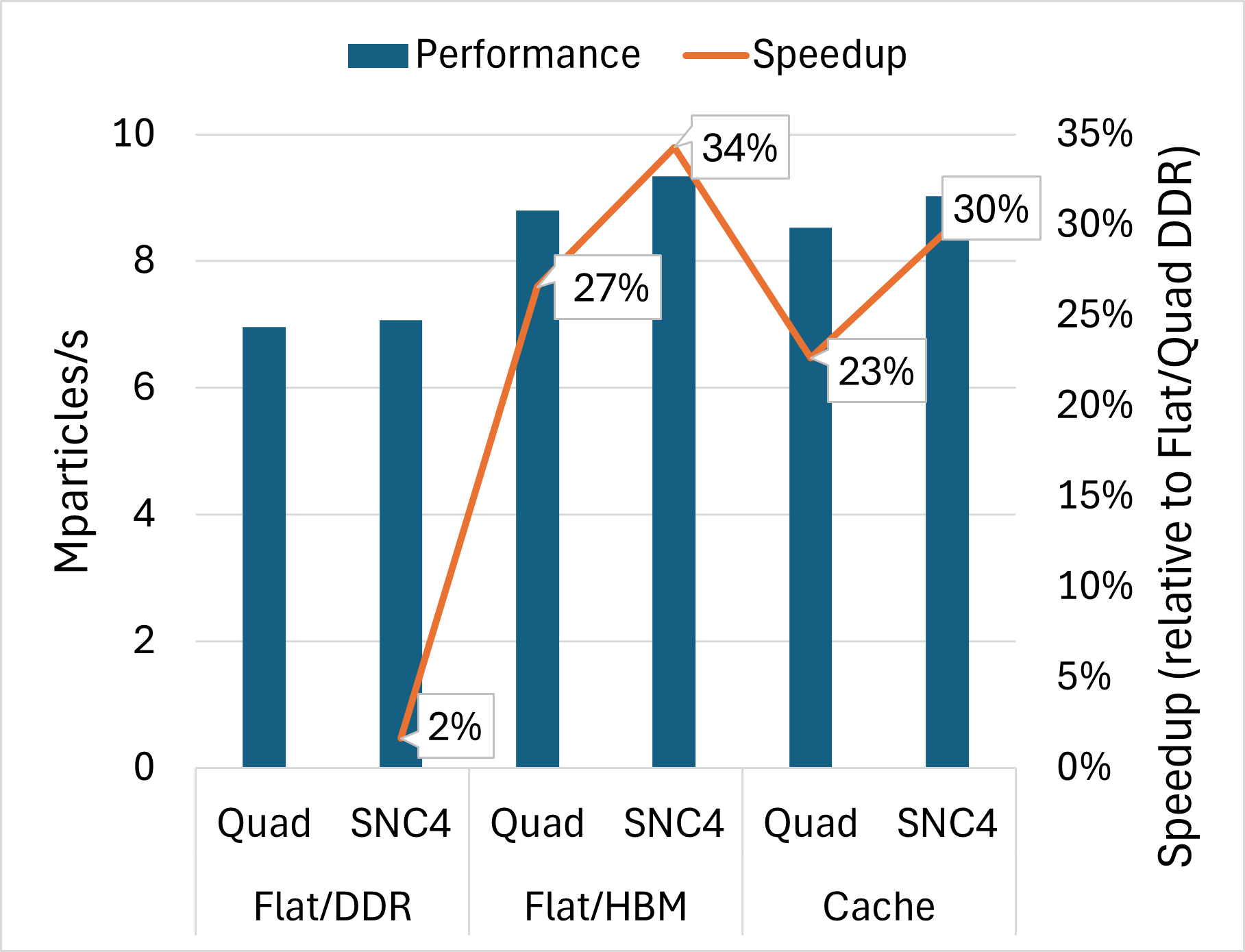}}
\caption{HACC performance across different modes, with speedup relative to Quad-Flat/DDR configuration.}
\label{fig:hacc_single_node}
\end{figure}

We begin by analyzing the single-node performance of HACC across different memory systems and configurations, using \(n_g = 912\) and an MPI geometry of \(6 \times 4 \times 4\). As shown in Figure~\ref{fig:hacc_single_node}, HACC performance strongly correlates with memory bandwidth, as outlined in Table~\ref{tab:stream_bw}. This is due to the memory bandwidth requirements of the TreeBuild, Poisson Solver, and ResortParticles components running on the CPU. For these components, Flat/HBM provides speedups ranging from 1.5$\times$ to 2$\times$ compared to Flat/DDR. Among the configurations tested, SNC4-Flat/HBM achieves the highest overall HACC performance.

\subsubsection{Weak Scaling Performance}

Next, we compare the weak scaling performance and efficiency on HBM and DDR memory systems. The weak scaling tests are conducted using node counts of 128, 1024, and 8192. Since the simulation domain is a cube, increasing the node count by a factor of eight scales the grid size (\(n_g\)) by a factor of two along each dimension, leading to an eightfold increase in the total number of particles. The grid size (\(n_g\)) and MPI geometry for each node count are shown in Table~\ref{tab:hacc_config}.
\begin{table}[htbp]
    \centering
    \caption{HACC configurations.}
    \renewcommand{\arraystretch}{1.3}
    \begin{tabular}{ccc}
        \toprule
        \textbf{Node Count} & \textbf{Grid Size (\(n_g\))} & \textbf{MPI Geometry} \\ 
        \midrule
         128 &  4608 & $32 \times 24 \times 16$ \\
         1024 &  9216 & $64 \times 48 \times 32$ \\
         8192 &  18432 & $128 \times 96 \times 64$ \\
        \bottomrule
    \end{tabular}
    \label{tab:hacc_config}
\end{table}

\begin{figure}[htbp]
\centerline{\includegraphics[width=0.75\linewidth]{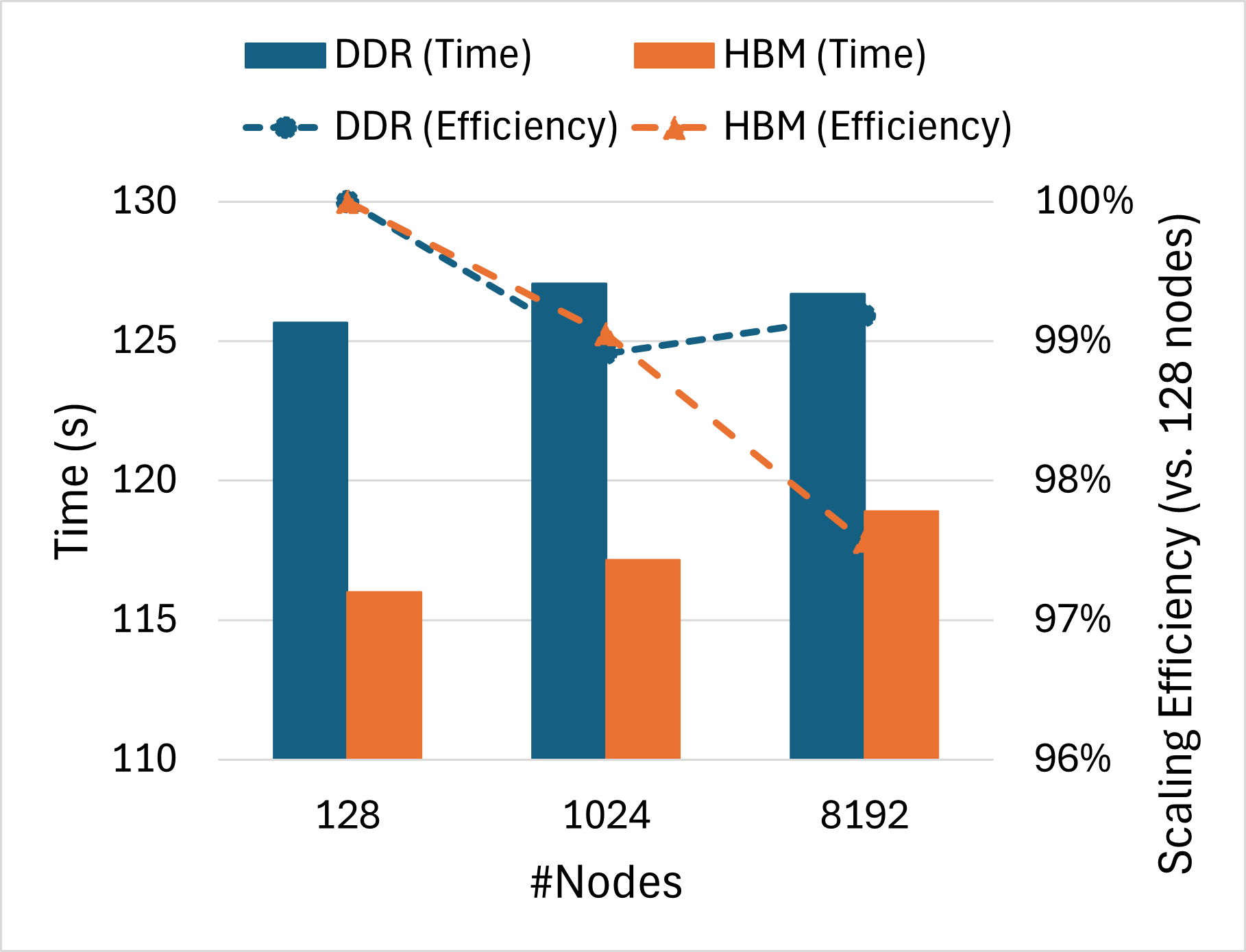}}
\caption{Weak scaling performance and efficiency of HACC on Quad-Flat/HBM and Quad-Flat/DDR. Bars represent execution time, with lower values indicating better performance, while lines show scaling efficiency.}
\label{fig:hacc_scaling}
\end{figure}

Figure~\ref{fig:hacc_scaling} shows that execution time, represented by the bars on the left y-axis, is consistently lower for HBM compared to DDR across all tested node counts. As the node count increases, HBM continues to exhibit faster execution times than DDR. However, DDR demonstrates better scaling efficiency, particularly at larger node counts.

The scaling efficiency results, shown as lines on the right y-axis, indicate that both memory systems maintain high efficiency as the node count increases. At 1024 nodes, both HBM and DDR achieve approximately 99\% scaling efficiency (relative to 128 nodes). As the system scales to 8192 nodes, DDR sustains above 99\% efficiency, while HBM efficiency decreases to 97\%. This divergence suggests that the longer computation time of DDR allows for more effective overlap with communication tasks, enabling higher efficiency at larger scales. Nonetheless, both HBM and DDR exhibit excellent scaling efficiency across the tested configurations.

\subsection{QMCPACK}

QMCPACK is a modern, high-performance Quantum Monte Carlo (QMC) simulation code designed for electronic structure calculations of molecules, nanomaterials, and solid-state systems~\cite{qmcpack_2018}. It explicitly solves the many-body Schrödinger equation using stochastic sampling. The computationally intensive routines in QMCPACK primarily involve dense matrix and vector operations, with optimized implementations leveraging BLAS for high vectorization efficiency. The computational cost of Fixed-Node Diffusion Monte Carlo (DMC) scales as \(O(N^3)\) to \(O(N^4)\), where \(N\) is the number of electrons, while maintaining near-linear scaling with the number of computing nodes. This strong scalability makes QMCPACK well-suited for leadership-class supercomputers, such as the Aurora exascale system.

The QMCPACK code employs a hybrid parallelization strategy: i) MPI across NUMA nodes to reduce memory footprint; ii) OpenMP threading to parallelize Monte Carlo (MC) samples for independent progress; and iii) OpenMP offloading to accelerate fine-grained data parallelism in sample-level computations~\cite{hipar_2022}. The weak-scaling efficiency of levels (i) and (ii) remains high due to the infrequent synchronization through MPI collectives and walker rebalancing, both of which are largely independent of the memory mode. Therefore, our analysis focuses on single-node performance.

\subsubsection{Single-Node Performance}
\begin{figure}[htbp]
    \centering
    \begin{subfigure}[b]{0.75\linewidth}
        \centering
        \includegraphics[width=\linewidth]{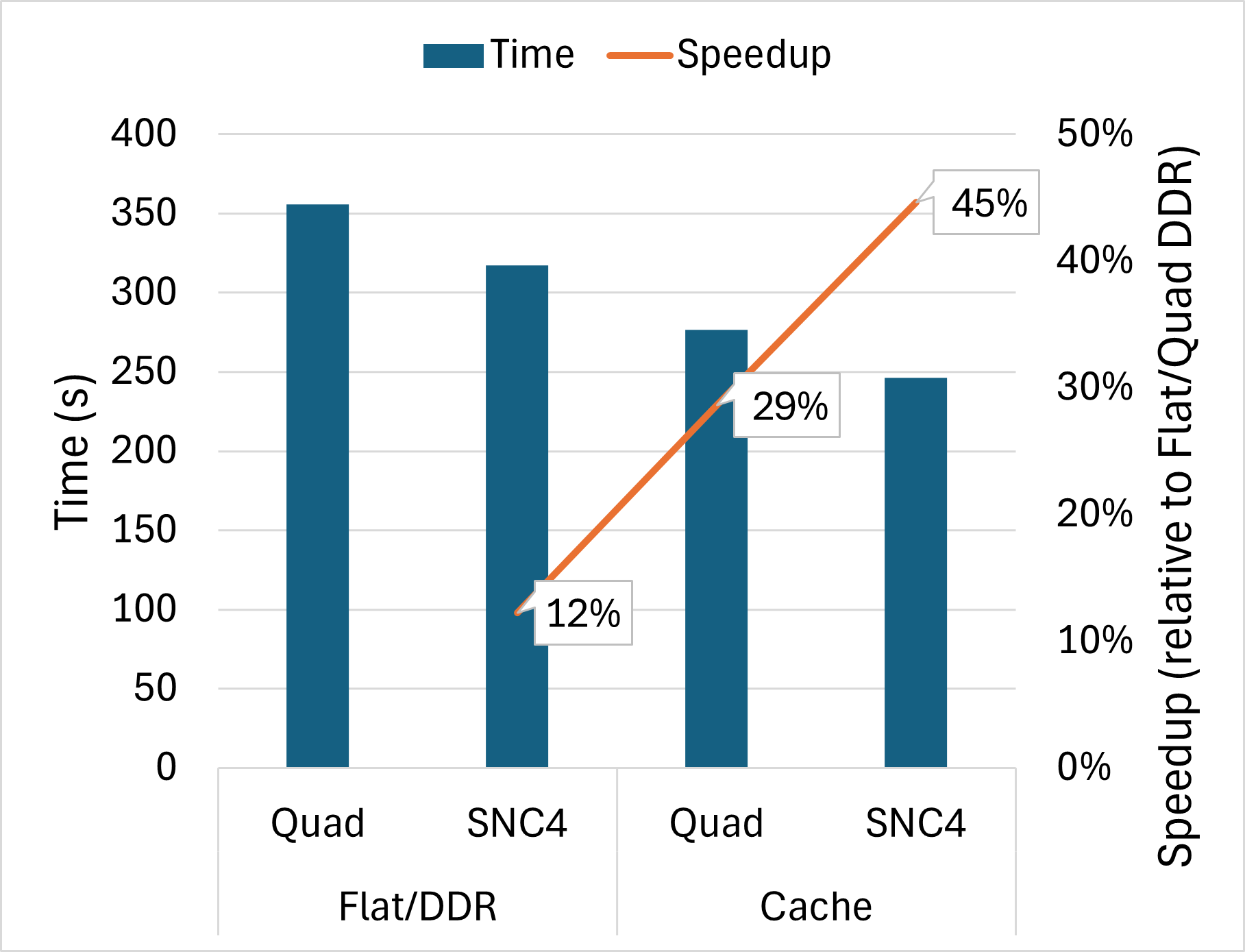}
        \caption{CPU-only}
        \label{fig:qmcpack_cpu}
    \end{subfigure}
    \begin{subfigure}[b]{0.75\linewidth}
        \centering
        \includegraphics[width=\linewidth]{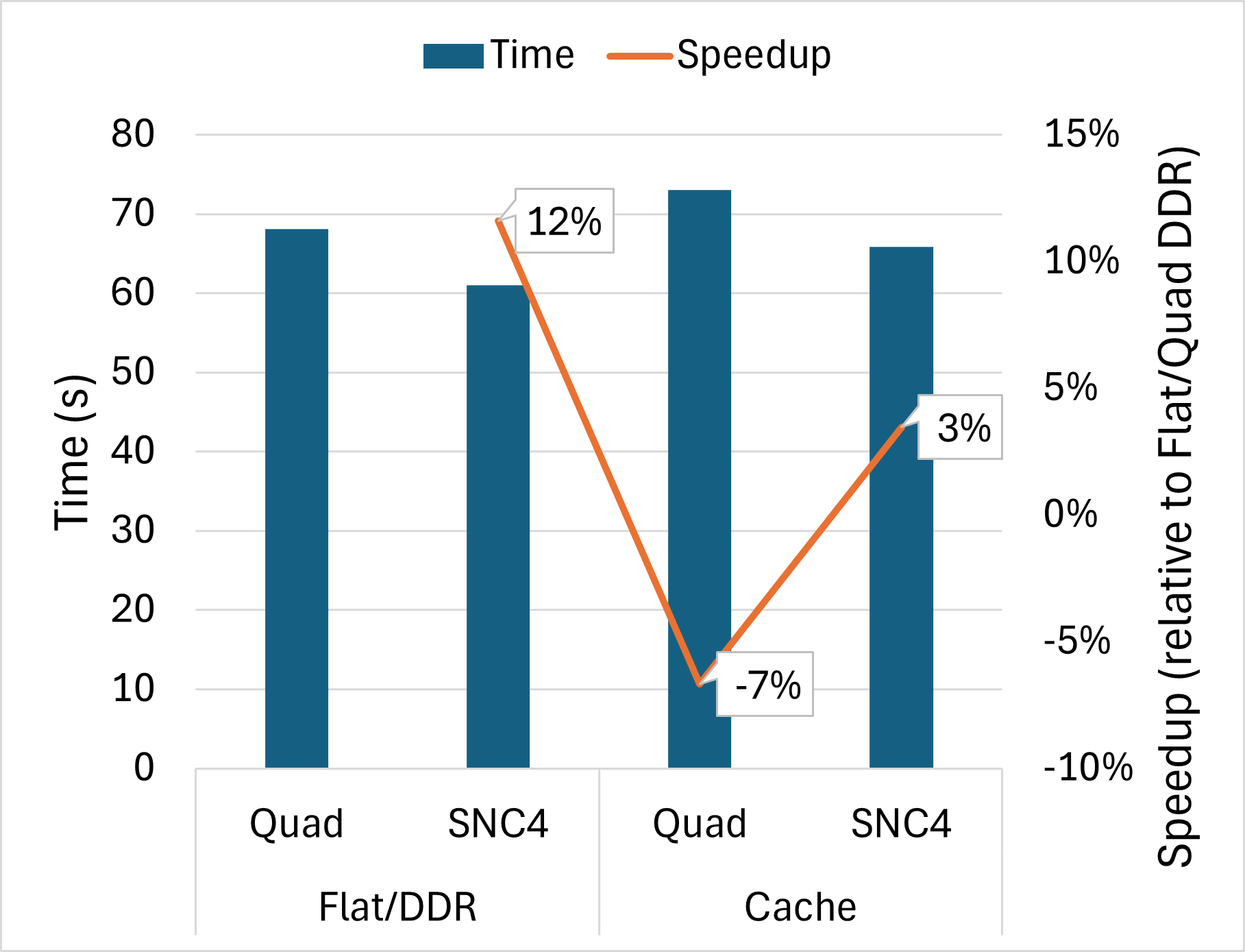}
        \caption{Hybrid CPU/GPU}
        \label{fig:qmcpack_hybrid}
    \end{subfigure}
    \caption{QMCPACK performance across different memory modes for CPU-only and hybrid CPU/GPU computations, with speedup relative to the Quad-Flat/DDR configuration.}
    \label{fig:qmcpack_single_node}
\end{figure}

Figure~\ref{fig:qmcpack_single_node} shows the timing of a DMC calculation for a 512-atom crystalline B1 NiO with $N=6144$ electrons~\cite{nio_2017}. This workload is representative of the scale of the Aurora exascale system, and its performance characteristics are largely transferable to production simulations. We use \texttt{PPN=8}, specifying 8 processes per node, and 12 OpenMP threads for CPU-only runs (Figure~\ref{fig:qmcpack_cpu}), and \texttt{PPN=12}, with one rank per GPU tile and 8 OpenMP threads for hybrid CPU/GPU runs (Figure~\ref{fig:qmcpack_hybrid}). In both cases, we evaluate the performance using Flat/DDR and Cache memory modes, as the memory footprint exceeds the 64 GB HBM capacity per socket.

In QMCPACK hybrid CPU/GPU runs with \texttt{PPN=12} and 8 OpenMP threads, we ensure an even distribution of cores across all quadrants in SNC4 mode. Most processes are pinned to a single quadrant, while some span two quadrants to balance the core distribution. This is shown in Table~\ref{tab:qmcpack_hybrid_pinning} for a single socket.
\begin{table}[htbp]
    \centering
    \caption{Process pinning and core assignment for a single socket in QMCPACK hybrid CPU/GPU runs.}
    \renewcommand{\arraystretch}{1.3}
    \begin{tabular}{ccc}
    \hline
    \textbf{Process} & \textbf{Cores Assigned} & \textbf{CPU Quadrant} \\
    \toprule
    0  & 1-8   & 0   \\
    1  & 9-16  & 0, 1 \\
    2  & 17-24 & 1   \\
    3  & 27-34 & 2   \\
    4  & 35-42 & 2, 3 \\
    5  & 43-50 & 3   \\
    \bottomrule
    \end{tabular}
    \label{tab:qmcpack_hybrid_pinning}
\end{table}

The critical computations in the workload include (i) calculating single-particle orbitals (SPOs), (ii) inverting matrices of size $N/2 \times N/2$ in double precision, and (iii) updating inverse matrices using a rank-k update algorithm based on the Woodbury identity in single precision. The bulk of these computations are handled by the highly optimized oneMKL library. All computational paths can run on either CPU or GPU. We select the optimal execution path which consists of an approximately 50/50 split between CPU and GPU.

The arithmetic intensity varies, ranging from memory bandwidth-bound SPO calculations to FLOP-bound GEMMs in (ii) and (iii). In CPU-only runs, the total available DDR bandwidth of a socket is a key performance-limiting factor, as evidenced by the 29\% performance improvement in Quad-Cache mode over Quad-Flat/DDR mode in Figure~\ref{fig:qmcpack_cpu}. We attribute the improvement in SNC4 mode, for both Flat/DDR and Cache configurations, to the lower latency of SNC4 compared to Quad mode. Offloading SPO calculations to GPUs significantly reduces the bandwidth demands on CPUs, whereas Cache mode does not offer the same benefit. The particle-by-particle update method, aimed at maximizing MC sampling efficiency, results in the invocation of many small kernels and frequent small data transfers to the GPU. As shown in Figure~\ref{fig:qmcpack_hybrid}, these factors contribute to the workload sensitivity to memory systems, particularly latency, which makes Flat/DDR perform better than Cache mode in hybrid CPU/GPU runs due to the lower latency of DDR for small and frequent memory requests.

\subsection{BFS}

Breadth-First Search (BFS) is a fundamental graph traversal algorithm widely used in high-performance computing applications. It is also a core component of the Graph500 benchmark~\cite{murphy2010graph500}, which evaluates the performance of supercomputing systems on large-scale graph problems. BFS is characterized by intensive irregular memory access patterns and low computational intensity. For this study, we use an optimized CPU-only BFS implementation from the RIKEN Graph500-BFS repository\footnote{\url{https://github.com/RIKEN-RCCS/Graph500-BFS}}, which incorporates advanced techniques to improve parallelism and scalability on high-performance systems. These techniques include forest pruning, group reordering, multilevel bitmap compression, and adaptive parameter tuning, all of which improve memory efficiency and communication performance~\cite{arai2024bfs}.

In the Graph500 benchmark, the graph consists of \(2^{\texttt{SCALE}}\) vertices and \(16 \times 2^{\texttt{SCALE}}\) edges, where \texttt{SCALE} is a positive integer parameter. BFS performance is measured in TEPS (Traversed Edges Per Second), which is defined as the number of edges traversed in the connected component containing a search key, divided by the processing time in seconds. The benchmark reports the harmonic mean of the TEPS values from 64 search keys, where each search key is a randomly selected vertex in the graph.

\subsubsection{Single-Node Performance}

We run BFS on a single node with the parameters \texttt{-A -C -n 64}, where \texttt{-A} enables adaptive parameter tuning, \texttt{-C} enables forest pruning, and \texttt{-n 64} sets the number of search keys to 64. Additionally, we configure the system to use \texttt{PPN=96}, specifying 96 processes per node.

\begin{figure}[htbp]
\centerline{\includegraphics[width=0.75\linewidth]{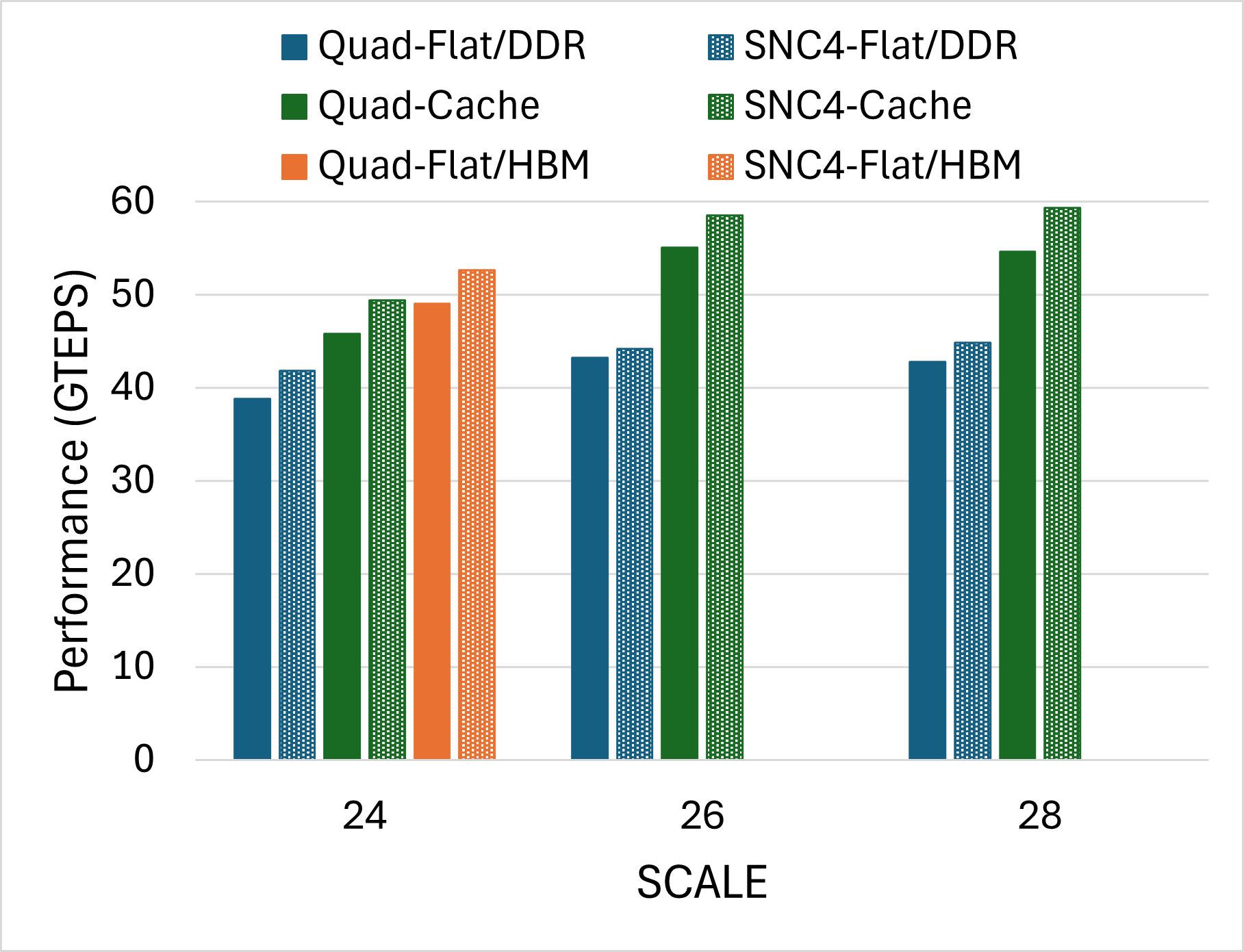}}
\caption{BFS performance across different graph sizes (\texttt{SCALE}) and memory configurations.}
\label{fig:bfs_single_node}
\end{figure}

Figure~\ref{fig:bfs_single_node} presents BFS performance for different graph sizes across various memory and clustering modes. At the smaller graph size (\texttt{SCALE=24}), where the graph fits entirely within HBM, performance strongly correlates with memory bandwidth, as outlined in Table~\ref{tab:stream_bw}. SNC4-Flat/HBM achieves the highest performance due to its higher bandwidth.

As the graph size increases, BFS performance improves due to enhanced parallelism, amortization of fixed overheads, and more efficient memory access patterns. For larger graphs that exceed HBM capacity (\texttt{SCALE=26} and \texttt{SCALE=28}), SNC4-Cache delivers the best performance, followed by Quad-Cache. Cache mode effectively utilizes HBM while leveraging the larger DDR memory to accommodate the full graph. Thus, SNC4-Flat/HBM achieves the best performance for graphs that fit within HBM, while SNC4-Cache becomes the most efficient configuration for larger graphs exceeding HBM capacity.

\subsubsection{Weak Scaling Performance}

\begin{figure}[htbp]
\centerline{\includegraphics[width=0.75\linewidth]{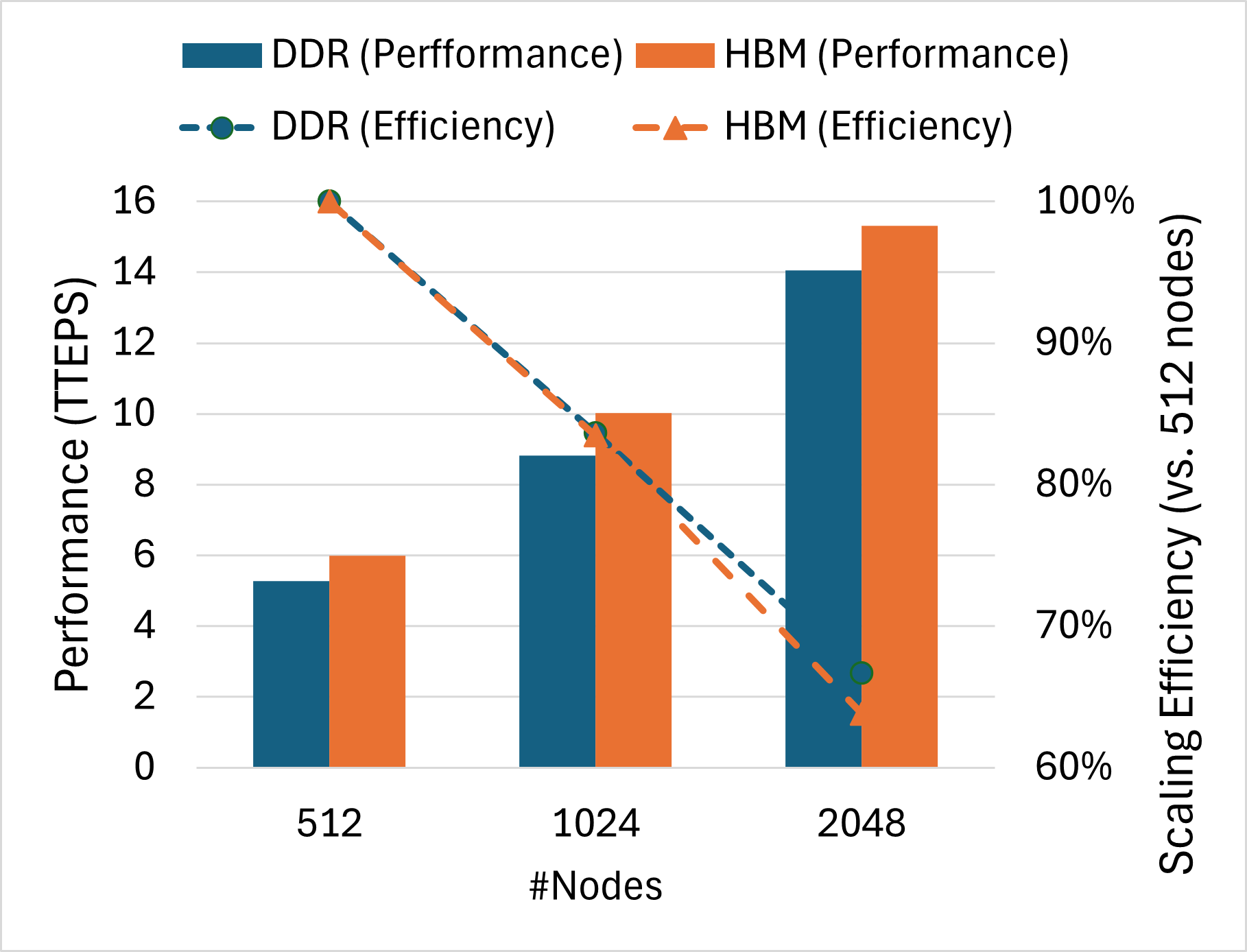}}
\caption{Weak scaling performance and efficiency of BFS on Quad-Flat/HBM and Quad-Flat/DDR. Bars represent performance (TEPS), where higher values indicate better performance, while lines show scaling efficiency.}
\label{fig:bfs_scaling}
\end{figure}

Figure~\ref{fig:bfs_scaling} shows the weak scaling performance and efficiency of BFS on HBM and DDR memory systems. In these experiments, we use 32 processes per node (\texttt{PPN=32}). The tests start with \texttt{SCALE=35} at 512 nodes, and \texttt{SCALE} is incremented by one for each doubling of the node count, which also results in doubling the graph size. Across all node counts, HBM achieves approximately 10\% higher performance than DDR. However, consistent with other applications, DDR demonstrates slightly better scaling efficiency compared to HBM.

\section{Related Work} \label{sec:relatedwork}

HBM has been widely studied in HPC architectures, particularly in comparison to traditional DDR-based memory systems. Previous research highlights key trade-offs in bandwidth scalability, memory latency, and application performance.

McCalpin~\cite{mccalpin2023bandwidth} examines sustained memory bandwidth limitations in Intel Xeon Max CPUs, showing that real-world performance is constrained by memory concurrency despite HBM high theoretical bandwidth. Fukazawa and Takahashi~\cite{fukazawa2024} evaluate Xeon Max CPUs and demonstrate that HBM benefits vary based on workload characteristics. Siegmann et al.~\cite{siegmann2024} compare HBM and DDR performance in scientific applications, identifying cases where DDR remains competitive.

Hybrid memory architectures have also been widely explored, particularly in Intel Knights Landing (KNL) processors, which, like Aurora CPUs, support multiple memory modes. Peng et al.\cite{peng2017} show that application performance is dependent on memory mode selection, while Ramos and Hoefler\cite{ramos2017} analyze memory efficiency in manycore systems. Salehian and Yan~\cite{salehian2017} evaluate HBM impact on scientific workloads, and Yount and Duran~\cite{yount2016} investigate its effectiveness as a high-bandwidth cache for stencil computations.

While prior studies provide valuable insights into HBM and DDR trade-offs, our work presents the first in-depth evaluation of these factors on Aurora, examining their implications for system efficiency and HPC workload performance.

\section{Discussion and Conclusion} \label{sec:discussion}

Memory systems such as DDR and HBM differ not only in bandwidth but also in latency and capacity, each of which impacts system performance in unique ways. HBM offers superior bandwidth, DDR exhibits lower latency for low memory access loads, and DDR provides larger memory capacity.

When evaluating memory systems and their configurations, it is crucial to consider how these properties affect various system performance metrics. For example, the performance benefits of higher bandwidth extend beyond CPU memory bandwidth-bound kernels, as increased bandwidth also influences other system metrics, such as PCIe transfers. Since each data transfer between the CPU and a PCIe device requires a corresponding memory read or write operation, memory characteristics can impact factors such as CPU-GPU PCIe bandwidth and MPI bandwidth.

In addition to memory systems, the choice of memory and clustering modes can affect performance differently depending on the application. For instance, Flat memory mode typically offers higher bandwidth compared to Cache mode, but it requires the application to explicitly manage NUMA node affinities. In contrast, Cache mode allows applications to leverage HBM while also benefiting from the larger DDR memory capacity, without requiring any modifications. Furthermore, SNC4 clustering mode often outperforms Quad mode by enhancing memory locality, reducing latency, and increasing bandwidth. However, its effectiveness relies on careful management of memory and processor affinities to avoid potential performance degradation. In contrast, Quad mode provides greater flexibility, particularly in complex configurations, such as when multiple GPUs share a quadrant or in cases of asymmetric NIC connectivity between quadrants, as seen in the Aurora compute node.

There is no one-size-fits-all configuration, but the following observations can help guide applications in selecting the best memory systems and configurations:
\begin{itemize}
\item For CPU-only applications or hybrid CPU/GPU applications with CPU memory bandwidth-bound kernels, performance improves with higher bandwidth. In these cases, HBM outperforms DDR, Flat/HBM memory mode outperforms Cache mode, and SNC4 clustering mode outperforms Quad mode, as demonstrated in HACC and CPU-only QMCPACK and BFS.
\item Applications heavily dependent on CPU-GPU PCIe bandwidth should prioritize Quad clustering mode and ensure memory allocations target HBM, whether in Flat or Cache memory mode, for optimal performance.
\item Applications sensitive to memory latency, such as those with small and frequent memory requests or GPU data transfers, perform better with DDR due to its lower latency in these workloads, as shown in hybrid QMCPACK.
\item While SNC4 clustering mode generally improves memory locality, reducing latency and increasing bandwidth, it has limitations. These include decomposition imbalances (as seen in hybrid QMCPACK with \texttt{PPN=12}) and reduced memory bandwidth and capacity per quadrant (16 GB per quadrant compared to 64 GB in Quad mode). SNC4 mode is most effective when the number of processes exceeds the number of quadrants, but it requires careful management of memory and processor affinities to prevent performance degradation. In contrast, Quad mode offers greater overall flexibility, with fewer performance penalties in complex configurations.
\end{itemize}

\section*{Acknowledgment}
This research used resources of the Argonne Leadership Computing Facility, a U.S. Department of Energy (DOE) Office of Science user facility at Argonne National Laboratory and is based on research supported by the U.S. DOE Office of Science-Advanced Scientific Computing Research Program, under Contract No. DE-AC02-06CH11357.


\bibliographystyle{IEEEtran.bst}
\bibliography{ref}

\begin{thebibliography}{10}
\providecommand{\url}[1]{#1}
\csname url@samestyle\endcsname
\providecommand{\newblock}{\relax}
\providecommand{\bibinfo}[2]{#2}
\providecommand{\BIBentrySTDinterwordspacing}{\spaceskip=0pt\relax}
\providecommand{\BIBentryALTinterwordstretchfactor}{4}
\providecommand{\BIBentryALTinterwordspacing}{\spaceskip=\fontdimen2\font plus
\BIBentryALTinterwordstretchfactor\fontdimen3\font minus
  \fontdimen4\font\relax}
\providecommand{\BIBforeignlanguage}[2]{{%
\expandafter\ifx\csname l@#1\endcsname\relax
\typeout{** WARNING: IEEEtran.bst: No hyphenation pattern has been}%
\typeout{** loaded for the language `#1'. Using the pattern for}%
\typeout{** the default language instead.}%
\else
\language=\csname l@#1\endcsname
\fi
#2}}
\providecommand{\BIBdecl}{\relax}
\BIBdecl

\bibitem{aurora}
\BIBentryALTinterwordspacing
\emph{{Aurora}}. [Online]. Available: \url{https://www.alcf.anl.gov/aurora}
\BIBentrySTDinterwordspacing

\bibitem{applencourt2024ponte}
T.~Applencourt, A.~Sadawarte, S.~Muralidharan, C.~Bertoni, J.~Kwack, Y.~Luo,
  E.~Rangel, J.~Tramm, Y.~Ghadar, A.~Tamerus \emph{et~al.}, ``{Ponte Vecchio}
  across the {Atlantic}: Single-node benchmarking of two {Intel GPU} systems,''
  in \emph{IEEE International Workshop on Performance Modeling, Benchmarking
  and Simulation of High Performance Computer Systems}, 2024.

\bibitem{mccalpin2023bandwidth}
J.~D. McCalpin, ``Bandwidth limits in the {Intel Xeon Max} ({Sapphire Rapids}
  with {HBM}) processors,'' in \emph{International Conference on High
  Performance Computing}.\hskip 1em plus 0.5em minus 0.4em\relax Springer,
  2023, pp. 403--413.

\bibitem{fukazawa2024}
\BIBentryALTinterwordspacing
K.~Fukazawa and R.~Takahashi, ``Performance evaluation of the fourth-generation
  {Xeon} with different memory characteristics,'' in \emph{Proceedings of the
  International Conference on High Performance Computing in Asia-Pacific Region
  Workshops}, ser. HPCAsia '24 Workshops.\hskip 1em plus 0.5em minus
  0.4em\relax New York, NY, USA: Association for Computing Machinery, 2024, p.
  55–62. [Online]. Available: \url{https://doi.org/10.1145/3636480.3637218}
\BIBentrySTDinterwordspacing

\bibitem{siegmann2024}
E.~Siegmann, R.~J. Harrison, D.~Carlson, S.~Chheda, A.~Curtis, F.~Coskun,
  R.~Gonzalez, D.~Wood, and N.~A. Simakov, ``First impressions of the {Sapphire
  Rapids} processor with {HBM} for scientific workloads,'' \emph{SN Computer
  Science}, vol.~5, no.~5, p. 623, 2024.

\bibitem{intel_xeon_max_guide}
\BIBentryALTinterwordspacing
\emph{{Intel {Xeon CPU Max} Series Configuration and Tuning Guide}}, 2023.
  [Online]. Available:
  \url{https://www.intel.com/content/www/us/en/content-details/787743/intel-xeon-cpu-max-series-configuration-and-tuning-guide.html}
\BIBentrySTDinterwordspacing

\bibitem{fake_numa}
\BIBentryALTinterwordspacing
\emph{{Fake NUMA For CPUSets}}. [Online]. Available:
  \url{https://docs.kernel.org/arch/x86/x86_64/fake-numa-for-cpusets.html}
\BIBentrySTDinterwordspacing

\bibitem{mccalpin1995memory}
J.~D. McCalpin \emph{et~al.}, ``Memory bandwidth and machine balance in current
  high performance computers,'' \emph{IEEE computer society technical committee
  on computer architecture (TCCA) newsletter}, vol.~2, no. 19-25, 1995.

\bibitem{osubenchmark}
\BIBentryALTinterwordspacing
D.~K. Panda \emph{et~al.}, ``{OSU} micro-benchmarks.'' [Online]. Available:
  \url{http://mvapich.cse.ohio-state.edu/benchmarks/}
\BIBentrySTDinterwordspacing

\bibitem{habib2016hacc}
S.~Habib, A.~Pope, H.~Finkel, N.~Frontiere, K.~Heitmann, D.~Daniel, P.~Fasel,
  V.~Morozov, G.~Zagaris, T.~Peterka \emph{et~al.}, ``{HACC}: Simulating sky
  surveys on state-of-the-art supercomputing architectures,'' \emph{New
  Astronomy}, vol.~42, pp. 49--65, 2016.

\bibitem{qmcpack_2018}
\BIBentryALTinterwordspacing
J.~K. et~al., ``{QMCPACK}: an open source ab initio quantum {Monte Carlo}
  package for the electronic structure of atoms, molecules and solids,''
  \emph{Journal of Physics: Condensed Matter}, vol.~30, no.~19, p. 195901, apr
  2018. [Online]. Available: \url{https://dx.doi.org/10.1088/1361-648X/aab9c3}
\BIBentrySTDinterwordspacing

\bibitem{hipar_2022}
Y.~Luo, P.~Doak, and P.~Kent, ``A high-performance design for hierarchical
  parallelism in the {QMCPACK Monte Carlo} code,'' in \emph{2022 IEEE/ACM
  International Workshop on Hierarchical Parallelism for Exascale Computing
  (HiPar)}, 2022, pp. 22--27.

\bibitem{nio_2017}
\BIBentryALTinterwordspacing
H.~Shin, Y.~Luo, P.~Ganesh, J.~Balachandran, J.~T. Krogel, P.~R.~C. Kent,
  A.~Benali, and O.~Heinonen, ``Electronic properties of doped and defective
  {NiO}: A quantum {Monte Carlo} study,'' \emph{Phys. Rev. Mater.}, vol.~1, p.
  073603, Dec 2017. [Online]. Available:
  \url{https://link.aps.org/doi/10.1103/PhysRevMaterials.1.073603}
\BIBentrySTDinterwordspacing

\bibitem{murphy2010graph500}
R.~C. Murphy, K.~B. Wheeler, B.~W. Barrett, and J.~A. Ang, ``Introducing the
  graph 500,'' \emph{Cray Users Group (CUG)}, vol.~19, no. 45-74, p.~22, 2010.

\bibitem{arai2024bfs}
J.~Arai, M.~Nakao, Y.~Inoue, K.~Teranishi, K.~Ueno, K.~Yamamura, M.~Sato, and
  K.~Fujisawa, ``Doubling graph traversal efficiency to 198 {TeraTEPS} on the
  supercomputer {Fugaku},'' in \emph{2024 SC24: International Conference for
  High Performance Computing, Networking, Storage and Analysis SC}.\hskip 1em
  plus 0.5em minus 0.4em\relax IEEE Computer Society, 2024, pp. 1616--1629.

\bibitem{peng2017}
I.~B. Peng, R.~Gioiosa, G.~Kestor, P.~Cicotti, E.~Laure, and S.~Markidis,
  ``Exploring the performance benefit of hybrid memory system on {HPC}
  environments,'' in \emph{2017 IEEE International Parallel and Distributed
  Processing Symposium Workshops (IPDPSW)}, 2017, pp. 683--692.

\bibitem{ramos2017}
S.~Ramos and T.~Hoefler, ``Capability models for manycore memory systems: A
  case-study with {Xeon Phi KNL},'' in \emph{2017 IEEE International Parallel
  and Distributed Processing Symposium (IPDPS)}, 2017, pp. 297--306.

\bibitem{salehian2017}
\BIBentryALTinterwordspacing
S.~Salehian and Y.~Yan, ``Evaluation of {Knight Landing} high bandwidth memory
  for {HPC} workloads,'' in \emph{Proceedings of the Seventh Workshop on
  Irregular Applications: Architectures and Algorithms}, ser. IA3'17.\hskip 1em
  plus 0.5em minus 0.4em\relax New York, NY, USA: Association for Computing
  Machinery, 2017. [Online]. Available:
  \url{https://doi.org/10.1145/3149704.3149766}
\BIBentrySTDinterwordspacing

\bibitem{yount2016}
C.~Yount and A.~Duran, ``Effective use of large high-bandwidth memory caches in
  {HPC} stencil computation via temporal wave-front tiling,'' in
  \emph{Proceedings of the 7th International Workshop on Performance Modeling,
  Benchmarking and Simulation of High Performance Computing Systems}, ser. PMBS
  '16.\hskip 1em plus 0.5em minus 0.4em\relax IEEE Press, 2016, p. 65–75.

\end{thebibliography}

\end{document}